\documentclass[aip,preprint]{revtex4-1}
\draft
\usepackage{graphicx}
\usepackage{bm}
\usepackage{color}

\begin{document}

\title{Surface transfer doping of hydrogen-terminated diamond probed by shallow nitrogen-vacancy centers}

\author{Taisuke Kageura}
\affiliation{Research Center for Materials Nanoarchitectonics, National Institute for Materials Science (NIMS), Tsukuba 305-0044, Japan}
\affiliation{National Institute of Advanced Industrial Science and Technology (AIST), Tosu 841-0052, Japan}

\author{Yosuke Sasama}
\affiliation{International Center for Young Scientists, National Institute for Materials Science (NIMS), Tsukuba 305-0044, Japan}

\author{Keisuke Yamada}
\affiliation{National Institutes for Quantum Science and Technology (QST), Takasaki 370-1292, Japan}

\author{Kosuke Kimura}
\affiliation{National Institutes for Quantum Science and Technology (QST), Takasaki 370-1292, Japan}
\affiliation{Graduate School of Science and Technology, Gunma University, Kiryu, 376-8515, Japan}

\author{\hspace{12mm}Shinobu Onoda}
\affiliation{National Institutes for Quantum Science and Technology (QST), Takasaki 370-1292, Japan}

\author{Yamaguchi Takahide}
\email[]{Corresponding author. E-mail address: yamaguchi.takahide@nims.go.jp}
\affiliation{Research Center for Materials Nanoarchitectonics, National Institute for Materials Science (NIMS), Tsukuba 305-0044, Japan}
\affiliation{University of Tsukuba, Tsukuba, 305-8571, Japan}

\begin{abstract}

The surface conductivity of hydrogen-terminated diamond is a topic of great interest from both scientific and technological perspectives. This is primarily due to the fact that the conductivity is exceptionally high without the need for substitutional doping, thus enabling a wide range of electronic applications. Although the conductivity is commonly explained by surface transfer doping due to air-borne surface acceptors, there remains uncertainty regarding the main determining factors that govern the degree of band bending and hole density, which are crucial for the design of electronic devices. Here, we elucidate the dominant factor influencing band bending by creating shallow nitrogen-vacancy (NV) centers beneath the hydrogen-terminated diamond surface through nitrogen ion implantation at varying fluences. We measured the photoluminescence and optically detected magnetic resonance (ODMR) of the NV centers, as well as the surface conductivity, as a function of the nitrogen implantation fluence. The disappearance of the conductivity with increasing nitrogen implantation fluence coincides with the appearance of photoluminescence and ODMR signals from negatively charged NV centers. This finding indicates that band bending is not exclusively determined by the work-function difference between diamond and the surface acceptor material, but by the finite density of surface acceptors. This work emphasizes the importance of distinguishing work-function-difference-limited band bending and surface-acceptor-density-limited band bending when modeling the surface transfer doping, and provides useful insights for the development of devices based on hydrogen-terminated diamond.
\end{abstract}

\keywords{Diamond, Transfer doping, NV center, Two-dimensional hole gas,\\ Hydrogen-terminated surface}

\maketitle
\renewcommand{\thesection}{\arabic{section}}
\renewcommand{\thesubsection}{\thesection.\arabic{subsection}}
\renewcommand{\thesubsubsection}{\thesubsection.\arabic{subsubsection}}

\section{Introduction}

Diamond has excellent properties as a wide bandgap semiconductor and quantum material and has potential applications in power electronics\cite{Gei18,Kaw22}, communication\cite{Kaw22}, computing\cite{Wra06,Chi13}, and sensing\cite{Sch14,Bar20}. An unusual property of diamond that could be used for such applications is its surface conductivity when it is terminated with hydrogen\cite{Gei18,Cra21,Kaw22}. P-type surface conductivity appears when hydrogen-terminated diamond is exposed to air even when the diamond is not intentionally doped, offering a unique solution to the problem of inefficient charge-carrier generation in diamond through substitutional doping\cite{Gei18,Cra21}. The surface conductivity can be basically explained by the surface transfer doping concept\cite{Mai00,Ris06J,Gei18,Cra21}. Here, atmospheric adsorbates, such as water and NO$_2$, on hydrogen-terminated surfaces act as acceptors. Electrons in the valence band of diamond are transferred to the acceptors, which induces band bending and generates holes below the diamond surface. Apart from atmospheric adsorbates, deposited oxides also act as acceptor materials\cite{Gei18,Cra21,Kaw22}.

The magnitude of the band bending induced by surface transfer doping and the corresponding areal density of holes and ionized acceptors are important information for designing devices based on hydrogen-terminated diamond. The hole density directly relates to the sheet resistance and current density. The ionized acceptors behave as scattering sources, and the hole mobility therefore depends on their density. It is generally assumed that the band bending due to electron transfer proceeds until the Fermi level of diamond aligns with that of the (isolated) surface acceptor material or, equivalently, the surface potential energy of diamond reaches the work-function difference between the diamond and surface acceptor material\cite{Mai00,Ris06J,Cra21}. For certain ranges of surface acceptor density and nitrogen concentration in diamond, however, the band bending due to electron transfer would not be as large as that\cite{Ris02,Pet12}. Such a surface transfer doping limited by the surface acceptor density and nitrogen concentration has not been investigated in detail.

In this study, we implanted nitrogen ions with ranging from $10^{11}$ to $10^{13}$ cm$^{-2}$ in the surface of diamond and investigated the implantation-fluence dependence of the surface conductivity of air-exposed hydrogen-terminated diamond. Furthermore, we carried out photoluminescence (PL) and optically detected magnetic resonance (ODMR) measurements on shallow nitrogen-vacancy (NV) centers created from the implanted nitrogen, which provided information on the potential energy and electric field near the diamond surface. The results of the experiments were compared with simulations based on the Schr\"odinger-Poisson equations under the assumption that the surface potential is constant or that the surface ionized acceptor density is constant irrespective of the nitrogen implantation fluence. The results suggest that band bending is not exclusively determined by the work-function difference between diamond and the surface acceptor material, but by the finite density of surface acceptors, which is approximately $10^{12}$ cm$^{-2}$ under our experimental conditions. This study demonstrates that shallow NV centers can be used to gain insights into surface transfer doping of hydrogen-terminated diamond.

\section{Methods}

\subsection{Experimental methods}

Two samples (Sample A and B) were made from high-purity single-crystal diamond plates with a nitrogen concentration below 5 ppb and boron concentration below 1 ppb. (See Appendix B for details of the sample preparation.)  Schematic diagrams of the samples are shown in Fig. 1. Shallow NV centers were created by nitrogen ($^{15}$N) ion implantation with an implantation energy of 10 keV and subsequent annealing at $1000$\r{}C. The implanted nitrogen atoms (and created NV centers) were distributed with a mean depth of ${\approx} 15$ nm according to a SRIM (Stopping and Range of Ions in Matter) simulation\cite{Zie10}. The surface of the samples was divided into four sections that had different nitrogen implantation fluences: $1\times10^{11}$, $2\times10^{11}$, $5\times10^{11}$, and $1\times10^{12}$ cm$^{-2}$ in Sample A, and $1\times10^{12}$, $2\times10^{12}$, $5\times10^{12}$, and $1\times10^{13}$ cm$^{-2}$ in Sample B. The diamond surface was hydrogen-terminated in a hydrogen plasma in a microwave-plasma-assisted chemical vapor deposition (CVD) chamber.

The PL imaging and ODMR measurements were performed using a home-built confocal microscope system \cite{Kag22}. A green laser with a wavelength of 532 nm was used for excitation, and fluorescence from the NV centers was detected with an avalanche photo diode through a 648-nm long-pass filter or guided to a spectrometer through a 561-nm long-pass filter. To obtain the cw-ODMR spectra, the fluorescence intensity was measured by continuously irradiating the sample with the green laser and sweeping the frequency of the applied microwaves. The surface conductivity was measured with a two-terminal method using two prober needles in contact with the diamond surface. (See Appendix C for details of the measurement setup.)

The charge state of an NV center can be negative (-1), neutral (0), or positive (+1) depending on the Fermi level $E_\mathrm{F}$ at its position. These different charge states can be distinguished by making PL and ODMR measurements. (See Appendix A for details.) The calculated NV$^{+}$/NV$^{0}$ and NV$^{0}$/NV$^{-}$ transition levels are at $E_\mathrm{NV}^{+/0} = 1.1$ eV and $E_\mathrm{NV}^{0/-} = 2.7$ eV above the valence band maximum\cite{Dea14}. This means that the NV center is in the NV$^{-}$ state for $E_\mathrm{F} {\textgreater} E_\mathrm{NV}^{0/-}$, NV$^{0}$ state for $E_\mathrm{NV}^{+/0} {\textless} E_\mathrm{F} {\textless} E_\mathrm{NV}^{0/-}$, and NV$^{+}$ state for $E_\mathrm{F} {\textless} E_\mathrm{NV}^{+/0}$. Therefore, once the charge state of the shallow NV centers is determined from PL and ODMR, it can be used to identify the energy range of the Fermi level at the positions of the NV centers. The electric field at the positions of the NV centers can also be estimated from the ODMR frequency.\cite{Dol11,Bro18}

\subsection{Modeling}

The band bending (the dependence of the potential $\phi$ on the depth $z$) in hydrogen-terminated diamond was calculated by solving the Schr\"odinger-Poisson equations\cite{Hau11,Gro12,Bro18}. (See Section S1 of the Supplementary Material for details.) The band bending is caused by surface transfer doping (capture of electrons by surface acceptors) and the resulting positive charges (positively charged donors and holes) generated in diamond. Nitrogen in diamond acts as a donor with a deep level 1.7 eV below the conduction band minimum. Therefore, the band bending depends significantly on the implanted nitrogen density as well as the background bulk nitrogen concentration.

Boundary conditions are necessary for solving the Poisson equation. The boundary condition $\frac{d\phi}{dz}(z{\to}{\infty}) = 0$ is used for deep inside diamond. Two different boundary conditions are used at the surface ($z = 0$). One is that the surface potential $\phi(0)$ (relative to deep inside the diamond) is constant irrespective of the implantation fluence. The other is that the surface electric field $\frac{d\phi}{dz}(z = 0)$ ($\phi'(0)$) is constant irrespective of the implantation fluence. These two conditions correspond to the following situations.

The constant $\phi(0)$ boundary condition corresponds to the situation where the band bending due to the electron transfer proceeds until the surface potential energy $-e\phi(0)$ reaches the work function difference between the diamond and surface acceptor material. This situation is the one of the original surface transfer doping model.\cite{Mai00} The constant $\phi(0)$ boundary condition assumes that the surface acceptor density is large enough to generate positive charge in diamond that can bend the band until $-e\phi(0)$ reaches the work function difference.

The constant $\phi'(0)$ boundary condition corresponds to the situation where the ionized acceptor (negative charge) density on diamond surface is constant. Here, $\phi'(0) = \frac{e}{\epsilon_\mathrm{S}}n_\mathrm{SA}^{-}$, where $n_\mathrm{SA}^{-}$ is the density of negatively charged surface acceptors. This boundary condition is applicable when the density of surface acceptors is not large enough. Even when the surface acceptors are fully ionized, the band does not bend until $-e\phi(0)$ reaches the work function difference.

The difference in the two boundary conditions may become clearer by considering band bending in an ideal pn junction of silicon, for example. Here, let us assume that the densities of donors in the n-type layer and acceptors in the p-type layer are both $10^{17}$ cm$^{-3}$ and that the n-type layer is thick enough. If the p-type layer is also thick enough, the band bends until the total band bending coincides with the difference (0.83 eV at 300 K) between the work functions of the n- and p-type layers\cite{Jen22}. The depletion region extends to 73 nm depth in both layers. However, if the p-type layer is as thin as 1 nm (i.e., the total areal density of acceptors is limited to be $10^{10}$ cm$^{-2}$), the depletion region in the n-type region (and p-type region) extends only to 1 nm because of charge neutrality, resulting in a band bending of only $1.5{\times}10^{-4}$ eV, which is much smaller than the work-function difference. The band bending is limited by the finite areal density of acceptors in this case. Note that, even in this case, the Fermi level is equal everywhere when equilibrium is achieved. However, the Fermi level in the p-type layer is close to the conduction band minimum as in the n-type layer. This Fermi level position is far from the Fermi level of a p-type layer that is isolated or before the junction is formed.

\section{Results and discussion}

\subsection{Results of the simulation}

\subsubsection{Constant $\phi(0)$ boundary condition}

Let us first show the results for the constant $\phi(0)$ boundary condition. Figure 2a shows the band bending for implantation fluences of $1\times10^{11}$ and $7\times10^{12}$ cm$^{-2}$. The bulk nitrogen and boron concentrations are assumed to be 5 and 1 ppb, respectively. The surface potential energy $-e\phi(0)$ relative to deep inside the diamond is assumed to be 3.8 eV. The electron affinity of hydrogen-terminated diamond is -1.3 eV\cite{Mai01} and the Fermi level of diamond with nitrogen and boron concentrations of 5 and 1 ppb is -1.7 eV, indicating that the work function of the diamond is 0.4 eV. Therefore, the surface potential energy of 3.8 eV means that the work function of the surface acceptor material is 4.2 eV. This value is nearly the same as the one assumed in the original surface transfer doping model\cite{Mai00}.

For an implantation fluence of $10^{11}$ cm$^{-2}$, bulk nitrogen donors are ionized from the surface to deep inside the diamond and the band bending is gradual due to the small concentration (5 ppb) of the positively charged donors. (See Figs. S1b and S1d for the depth profile of ionized impurities.) In contrast, the band bends steeply for an implantation fluence of $7\times10^{12}$ cm$^{-2}$ due to the large density of ionized implanted nitrogen. (See Figs. S1f and S1h for the depth profile of ionized impurities.) Figure 3b shows the electric field at the surface and the negative surface charge density as a function of implantation fluence. The electric field and negative charge density increase with increasing implantation fluence. The negative charge density reaches $8.6\times10^{12}$ cm$^{-2}$ for an implantation fluence of $1\times10^{13}$ cm$^{-2}$. This means that this constant $\phi(0)$ boundary condition (for an implantation fluence less than $1\times10^{13}$ cm$^{-2}$) is valid only when the surface acceptor density is larger than $8.6\times10^{12}$ cm$^{-2}$. 

The hole density is affected by the strong dependence of the band bending on the implantation fluence. As the implantation fluence increases, the confinement potential for the hole gas becomes steeper (Fig. 2b). This makes the quantized levels (the maximum energies of the valence subbands) depart from the Fermi level, which results in a decrease in hole density (Fig. 3c). Note that the hole density also depends strongly on $\phi(0)$. The hole density decreases substantially even for $\left[\mathrm{N}_\mathrm{imp}\right] = 0$ as $-e\phi(0)$ decreases from 3.8 to 3.6 eV (Fig. S2b).

The charge state of the shallow NV centers is determined by the position of the Fermi level relative to the NV$^{0}$/NV$^{-}$ and NV$^{+}$/NV$^{0}$ transition levels. Figures 2c and 2d show the depth profiles of the charge state of the shallow NV centers for implantation fluences $\left[\mathrm{N}_\mathrm{imp}\right]$ of $1\times10^{11}$ and $7\times10^{12}$ cm$^{-2}$. For $\left[\mathrm{N}_\mathrm{imp}\right] = 1\times10^{11}$ cm$^{-2}$, all the NV centers are positively charged because the Fermi level is below the NV$^{+}$/NV$^{0}$ transition level for the entire depth range in which the NV centers are distributed. (See also Figs. S1c and S1d.) For $\left[\mathrm{N}_\mathrm{imp}\right] = 7\times10^{12}$ cm$^{-2}$, the band bends rapidly and the Fermi level crosses the NV$^{+}$/NV$^{0}$ and NV$^{0}$/NV$^{-}$ levels within a narrow range of ${\approx} 20$ nm below the surface (Fig. S1g). The charge state varies from NV$^{+}$ to NV$^{0}$ and from NV$^{0}$ to NV$^{-}$ as the position of the NV center becomes deeper (Fig. S1h). Figure 3d shows the integrated density of NV$^{+}$, NV$^{0}$, and NV$^{-}$ as a function of implantation fluence. A relatively high implantation fluence of ${\approx} 5.5\times10^{12}$ cm$^{-2}$ is required to obtain negatively charged NV centers for this constant $\phi(0)$ boundary condition. This implantation fluence does not significantly depend on $\phi(0)$ (Fig. S2a) and is much larger than the fluence (${\approx} 2\times10^{12}$ cm$^{-2}$) at which the hole layer disappears (Fig. 3c).

\subsubsection{Constant $\phi'(0)$ boundary condition}

Next, let us turn to the results for the constant $\phi'(0)$ boundary condition. Figure 2e shows the band bending for implantation fluences of $1\times10^{11}$ and $1.5\times10^{12}$ cm$^{-2}$. The negative surface charge density $n_\mathrm{SA}^{-}$ is assumed to be $1\times10^{12}$ cm$^{-2}$, and this corresponds to $\phi'(0)$ of 0.317 MV/cm. In the case of $\left[\mathrm{N}_\mathrm{imp}\right] = 1\times10^{11}$ cm$^{-2}$, the negative surface charge is balanced mostly with the positive charge of holes generated near the surface, but also with the positive charge of the ionized bulk nitrogen and ionized implanted nitrogen. Bulk nitrogen is ionized from the surface to deep inside the diamond. (See Figs. S3b and S3d for the depth profile of ionized impurities.) Because the concentration of bulk nitrogen is low, the band bends gradually and the Fermi level is close to the valence band maximum at $z = 0$. In the case of $\left[\mathrm{N}_\mathrm{imp}\right] = 1.5\times10^{12}$ cm$^{-2}$, the negative surface charge is balanced mostly with the positive charge of the ionized implanted nitrogen (Figs. S3f and S3h). The electric field appears only just below the surface and the magnitude of the band bending is small. The Fermi level is far above the valence band maximum, and holes are not generated. Figure 3e shows the surface potential energy plotted as a function of implantation fluence. The surface potential energy rapidly decreases when the implantation fluence approaches the negative surface charge density. Figure 3g shows the hole density plotted as a function of implantation fluence. The hole density at $\left[\mathrm{N}_\mathrm{imp}\right] = 0$ is slightly lower than the negative surface charge density (by the amount of the space charge density caused by the ionized bulk nitrogen). The hole density decreases with increasing implantation fluence and goes to zero at an implantation fluence slightly lower than the negative surface charge density.

The implantation-fluence dependence of the charge state of the NV centers for the constant $\phi'(0)$ boundary condition (Fig. 3h) significantly differs from that for the constant $\phi(0)$ boundary condition. (Fig. 3d) The charge state of the NV centers changes steeply from NV$^{+}$ to NV$^{0}$ and from NV$^{0}$ to NV$^{-}$ in a small range of implantation fluences, which reflects the steep decrease in the surface potential shown in Fig. 3e. The charge states of all the NV centers change at nearly the same implantation fluence (Figs 2g and 2h), which is in contrast to the distribution of different charge states along the depth direction for the constant $\phi(0)$ boundary condition (Fig. 2d). It is also worth noting that the transition of the charge state of the NV centers and ceasing of hole generation occur at nearly the same implantation fluence, which is characteristic to the constant $\phi'(0)$ boundary condition.

\subsubsection{Which surface boundary condition is appropriate?}

Which boundary condition at the surface is appropriate for a given situation would depend on the surface acceptor density and the work function difference between the diamond and surface acceptor material. If the surface acceptor density is large, and/or if the work function difference is small, the constant $\phi(0)$ boundary condition is appropriate. In contrast, if the surface acceptor density is small, and/or if the work function difference is large, the $\phi'(0)$ constant boundary condition is appropriate. (The applicable boundary condition in the present study may depend on the nitrogen implantation fluence. See Section S2 of the Supplementary Material.) The acceptor density and work function difference depend on the acceptor material and the condition for the gas adsorption or oxide deposition on hydrogen-terminated diamond. The work function of the atmospheric adsorbed water layer depends on the pH and hydrogen concentration\cite{Mai00}. The surface acceptor density would also strongly depend on the environment where the diamond is placed. The acceptor density would be limited if molecules that do not act as acceptors adsorb and cover the diamond surface faster than the atmospheric acceptors.

The calculations of band bending in hydrogen-terminated diamond for different nitrogen implantation fluences in Refs. \cite{Hau11,Gro12} assume an electrolyte layer on the diamond and a constant potential $\phi(-\infty)$ in the bulk electrolyte, providing results similar to our calculation under the constant $\phi(0)$ boundary condition. This boundary condition (a constant $\phi(-\infty)$) is reasonable for the case of electrolyte gating\cite{Gro12} because the potential in the bulk electrolyte is maintained by applying a gate bias, but whether it is appropriate or not for the case of air-exposed hydrogen-terminated diamond\cite{Hau11} remains to be examined. The band-bending calculations in Refs. \cite{Pet12,Bro18} assume a fixed density of ionized adsorbed acceptors, but neglect the formation of subbands due to quantum confinement. The calculation in Ref. \cite{Bro18} assumes acceptor-type surface defect states as well as adsorbed acceptors. The Fermi-level-dependent densities of ionized nitrogen and NV centers are not included in the space charge density in the Poisson equation in Ref. \cite{Pet12}. As shown below, our experimental results are in better agreement with the calculation under the constant $\phi'(0)$ boundary condition than they are with the calculation under the constant $\phi(0)$ boundary condition.

\subsection{Experimental results and discussion}

The PL measurements were carried out on samples with different implantation-fluence regions before and after hydrogen plasma treatment. The PL intensity is plotted against implantation fluence in Fig. 4a. The intensity is the average over the 20 $\mu$m $\times$ 20 $\mu$m area of the PL images taken through a 648-nm long-pass filter. The diamond surface before the hydrogenation is oxygen-terminated because of the acid treatment. The PL intensity of the oxygen-terminated samples increases nearly linearly with implantation fluence. This observation suggests that the yield of the NV centers created by the ion implantation and subsequent vacuum annealing is independent of the implantation fluence, which is consonant with an earlier study in a low fluence range\cite{Pez10}. The PL intensity decreases after the hydrogen plasma treatment, and the decrease is more prominent at low fluences. Figure 4b shows the ratio of the PL intensity of the hydrogen-terminated surface to that of the oxygen-terminated surface.

The decrease in the PL intensity after hydrogenation can be attributed predominantly to two mechanisms. One is the transformation of NV centers to NVH defects due to hydrogen diffusion\cite{Sta12}, and the other is the stabilization of the NV$^{+}$ state due to band bending caused by the hydrogenation and surface transfer doping\cite{Hau11}. Note that the transition from NV$^{-}$ to NV$^{0}$ also causes a decrease in PL intensity in our setup because a 648-nm long-pass filter was used for the PL intensity measurements. For a fluence of $10^{13}$ cm$^{-2}$, the NV centers should primarily be in the NV$^{-}$ state, along with a smaller number in the NV$^{0}$ state, whereas the NV$^{+}$ state can be ruled out, even after the hydrogenation, because of the high concentration of nitrogen that acts as a donor. Therefore, the decrease in the PL intensity for $\left[\mathrm{N}_\mathrm{imp}\right] = 1\times10^{13}$ cm$^{-2}$ is mainly due to the formation of NVH defects. As the probability of the NVH formation is not expected to be higher for a lower fluence, the decrease in the PL intensity after hydrogenation in the low fluence range is attributed to stabilization of the NV$^{+}$ (and NV$^{0}$) state due to band bending.

PL spectra were measured (using a 561-nm long-pass filter) to assign NV$^{0}$ and NV$^{-}$ contributions to the PL (Fig. 5a). The 637-nm zero phonon peak and broad phonon side band, which are characteristic to NV$^{-}$, appear for $\left[\mathrm{N}_\mathrm{imp}\right] {\ge} 1\times10^{12}$ cm$^{-2}$, while the 637-nm zero phonon peak disappears and the broad-band emission is shifted to a shorter wavelength for $\left[\mathrm{N}_\mathrm{imp}\right] {\textless} 5\times10^{11}$ cm$^{-2}$. A similar effect of the nitrogen fluence on the PL spectra was reported in Ref. \cite{Gro14} for NV centers below the hydrogen-terminated surface created with 5-keV nitrogen implantation. To evaluate the NV$^{-}$/NV$^{0}$ population ratio, the measured PL spectra were fitted using NV$^{0}$ and NV$^{-}$ reference spectra. The NV$^{0}$ and NV$^{-}$ reference samples were prepared by controlling the density of donors (nitrogen) and acceptors (boron). The relative contributions of NV$^{0}$ and NV$^{-}$ centers to the PL were evaluated from a least-squares fit of the PL spectra with a linear combination of the reference spectra. (See Section S3 of the Supplementary Material for the details of the preparation of NV$^{0}$ and NV$^{-}$ reference spectra and fitting.) Then, the NV$^{+}$/NV$^{0}$/NV$^{-}$ population ratio was calculated as shown in Fig. 5b using the NV$^{0}$ and NV$^{-}$ contributions to PL and the PL-intensity ratio between the oxygen- and hydrogen-terminated surface. (See Section S4 in the Supplementary Material for details of the calculation.) The NV$^{+}$ state is dominant at low fluences, while the proportion of NV$^{-}$ centers increases for $\left[\mathrm{N}_\mathrm{imp}\right] {\ge} 1\times10^{12}$ cm$^{-2}$. The transition of the charge state was correlated with the implantation-fluence dependence of the surface conductivity (Fig. 5c). The conductivity decreases as the implantation fluence increases, and it nearly disappears when the implantation fluence reaches $(1-2)\times10^{12}$ cm$^{-2}$, at which the NV centers start to be negatively charged.

Figure 6 shows ODMR spectra obtained at 15 randomly selected spots in the PL images ($20$ $\mu$m $\times$ $20$ $\mu$m area) of different fluence regions (Fig. S7). No ODMR dips are visible for $\left[\mathrm{N}_\mathrm{imp}\right] = 2\times10^{11}$ and $5\times10^{11}$ cm$^{-2}$, while ODMR dips are visible in all the spectra for $\left[\mathrm{N}_\mathrm{imp}\right] {\ge} 1\times10^{12}$ cm$^{-2}$. This indicates the presence of the NV$^{-}$ centers for $\left[\mathrm{N}_\mathrm{imp}\right] {\ge} 1\times10^{12}$ cm$^{-2}$. Here, only two dips are visible in the spectra although PL signals from ensembles of NV centers were detected. This is because the magnetic field was aligned along the [001] axis of the diamond crystal to make the four possible N-V directions relative to the field direction equivalent (Fig. S9) and to enable a quantitative comparison of ODMR contrasts for different implantation fluences. Figures 7a and 7b show the implantation fluence dependence of the ODMR contrasts obtained by Lorentzian fits to the ODMR spectra (Fig. S8). The ODMR contrasts increase with increasing fluence, suggesting an increase in the NV$^{-}$/NV$^{0}$ ratio. This is consistent with a relative change in the NV$^{-}$ concentration obtained from the PL spectra (Fig. 5b). It is worth noting that Rabi oscillations were also observed for $\left[\mathrm{N}_\mathrm{imp}\right] {\ge} 1\times10^{12}$ cm$^{-2}$ (Fig. S10). Figures 7c and 7d show the implantation fluence dependence of the ODMR frequencies. The frequencies are independent of the implantation fluence within an experimental precision of ${\le} 0.3$ MHz. This indicates that the electric field at the position of the NV$^{-}$ centers does not substantially depend on implantation fluence, as discussed below. Note that the ODMR measurements were made at 15 individual fluorescence spots for $\left[\mathrm{N}_\mathrm{imp}\right] = 1\times10^{11}$ cm$^{-2}$. Dips are visible in the ODMR spectra of one of these spots (Fig. 6a and S8a). As individual fluorescence spots can be distinguished for $\left[\mathrm{N}_\mathrm{imp}\right] = 1\times10^{11}$ cm$^{-2}$  (Fig. S7a), the 15 spots cannot be regarded as randomly selected. If they were randomly selected in the whole image (not limited to the fluorescence spots), the probability of detecting ODMR would be close to zero, as is the case for $\left[\mathrm{N}_\mathrm{imp}\right] = 2\times10^{11}$ and $5\times10^{11}$ cm$^{-2}$.

Here, let us compare the experimental results with simulations. The observation that the NV$^{-}$ state appears for $\left[\mathrm{N}_\mathrm{imp}\right] {\ge} 1\times10^{12}$ cm$^{-2}$ is difficult to explain with the band-bending model with the constant $\phi(0)$ boundary condition. If we keep assuming that the surface potential energy $-e\phi(0)$ is 3.8 eV, the NV$^{0}$/NV$^{-}$ transition energy level $E_\mathrm{NV}^{0/-}$ must be assumed to be very low (${\approx} 0.6$ eV above the valence band maximum) to make the NV$^{-}$ state stable for $\left[\mathrm{N}_\mathrm{imp}\right] {\approx} 1\times10^{12}$ cm$^{-2}$. This transition energy level is substantially lower than 2.7-2.9 eV (above the valence band maximum) reported in the literature\cite{Dea14, Asl13}. If we use a value of 2.7 eV for $E_\mathrm{NV}^{0/-}$, $-e\phi(0)$ must be as low as ${\approx} 1.6$ eV to make the NV$^{-}$ state stable for $\left[\mathrm{N}_\mathrm{imp}\right] {\approx} 1\times10^{12}$ cm$^{-2}$. The low surface potential energy means a large (positive) electron affinity of diamond and/or a small work function of surface acceptors. As the surface potential energy decreases, the hole density decreases rapidly (Fig. S2b). No hole appears for the small surface potential energy of ${\approx} 1.6$ eV even at $\left[\mathrm{N}_\mathrm{imp}\right] = 0$, which is inconsistent with the experimentally observed conductivity. The simulation in Ref. \cite{Hau11} also indicates that NV centers created with an implantation energy of 10 keV are not negatively charged for $\left[\mathrm{N}_\mathrm{imp}\right] {\le} 3\times10^{12}$ cm$^{-2}$.

The band-bending model with the constant $\phi'(0)$ boundary condition can explain the observation of NV$^{-}$ centers for $\left[\mathrm{N}_\mathrm{imp}\right] {\approx} 1\times10^{12}$ cm$^{-2}$ if a negative surface charge density of ${\approx} 1\times10^{12}$ cm$^{-2}$ is assumed, as shown in Fig. 3h. The presence of the negative surface charge also leads to the generation of holes with a density of ${\approx} 1\times10^{12}$ cm$^{-2}$ for $\left[\mathrm{N}_\mathrm{imp}\right] = 0$ (Fig. 3g). As the hole mobility of hydrogen-terminated diamond exposed to air is ${\approx} 200$ cm$^{2}$V$^{-1}$s$^{-1}$ (Ref. \cite{Sas18}), the hole density of $1\times10^{12}$ cm$^{-2}$ corresponds to a conductivity of $3\times10^{-5}$ ${\Omega}^{-1}$, which is in reasonable agreement with those ($2.6\times10^{-5}$ ${\Omega}^{-1}$ and $1.7\times10^{-5}$ ${\Omega}^{-1}$) observed in the non-implanted region of the samples. It is worth noting that the value of the negative surface charge density is comparable to the one estimated from the mobility of field-effect transistors made of hydrogen-terminated diamond exposed to air and a hexagonal boron nitride gate insulator\cite{Sas20}. Furthermore, the model with the constant $\phi'(0)$ boundary condition is consistent with the finding that ceasing of hole conduction and an increase in the NV$^{-}$ ratio occur at a similar nitrogen implantation fluence.

Another experimental observation that is consistent with the constant $\phi'(0)$ boundary condition is the nitrogen-implantation-fluence dependence of the ODMR frequency. The positions of the dips in the ODMR spectra are independent of the implantation fluence within an experimental precision of ${\le} 0.3$ MHz. (Figs. 7c and 7d) When the constant $\phi'(0)$ boundary condition is used, the calculated average electric field at the positions of the NV$^{-}$ centers is smaller than $0.17$ MV/cm and is a decreasing function of implantation fluence for $\left[\mathrm{N}_\mathrm{imp}\right] {\ge} 1\times10^{12}$ cm$^{-2}$ (Fig. 7f). In this case, the shift in the ODMR frequency due to the variation in the electric field is less than 0.1 MHz (Figs. 7g and 7h; See Section S5 of the Supplementary Material for the details of the calculation.), which could not be resolved in our experiment. When the constant $\phi(0)$ boundary condition is used, however, the calculated average electric field at the positions of the NV$^{-}$ centers varies from ${\approx} 0.1$ to ${\approx} 0.9$ MV/cm with increasing implantation fluence (Fig. 7e). This increase in the electric field shifts the ODMR frequency by ${\approx} 2$ MHz (Figs. 7g and 7h), which could be detected experimentally. Thus, the nitrogen-implantation-fluence dependence of the ODMR frequency is also more consistent with the constant $\phi'(0)$ boundary condition than the constant $\phi(0)$ boundary condition.

The band-bending model with the constant $\phi'(0)$ boundary condition thus essentially explains the nitrogen-implantation-fluence dependences of the PL, ODMR, and conductivity. However, there is some discrepancy between the experiment and calculation. The calculated $\left[\mathrm{N}_\mathrm{imp}\right]$ dependence of NV$^{+}$/NV$^{0}$/NV$^{-}$ ratio has a sharp threshold near $\left[\mathrm{N}_\mathrm{imp}\right] = 1{\times}10^{12}$ cm$^{-2}$ (Fig. 3h), whereas the experimental transition from NV$^{+}$ to NV$^{-}$ is gradual (Fig. 5b).  The NV$^{-}$ state and hole conductivity appear to coexist at $\left[\mathrm{N}_\mathrm{imp}\right] = 1\times10^{12}$ cm$^{-2}$ (Figs. 5b and 5c), which is not explained by the calculation (Figs. 3g and 3h). The discrepancy may be caused by spatial inhomogeneities in the negative surface charge and implanted nitrogen, which are not considered in the calculation. For implantation fluences near the threshold, conductive non-fluorescent (NV$^{+}$) regions and insulating fluorescent (NV$^{-}$) regions may coexist within a length scale of the resolution of the optical microscope. When the average density of the negative surface charge (or implanted nitrogen) is $1{\times}10^{12}$ cm$^{-2}$, which corresponds to $1/(10$ nm$)^{2}$, it is likely that its local density varies on a length scale of order $10$ nm. Specifically, the negative surface charges may concentrate on the step edges of the hydrogen-terminated diamond (001) surface\cite{Gei21}, which could lead to coexistence of filamentary conductive regions and insulating fluorescent regions on such a length scale. The smaller ODMR contrast for $\left[\mathrm{N}_\mathrm{imp}\right] = 1\times10^{12}$ and $2\times10^{12}$ cm$^{-2}$ than for $\left[\mathrm{N}_\mathrm{imp}\right] = 5\times10^{12}$ and $1\times10^{13}$ cm$^{-2}$ (Figs. 7a and 7b) could be caused by the presence of intermediate regions where the NV centers are in the NV$^{0}$ state. The finite NV$^{0}$/NV$^{-}$ ratio for $\left[\mathrm{N}_\mathrm{imp}\right] = 1{\times}10^{13}$ cm$^{-2}$ may be caused by photo-induced conversion from NV$^{-}$ to NV$^{0}$ due to the laser light illumination\cite{Asl13}.

\section{Conclusions}

We have shown that shallow NV centers below the hydrogen-terminated diamond surface created with different nitrogen implantation fluences can provide novel information on the dominant factor that determines the band bending in surface transfer doping. In particular, we found that, with increasing nitrogen implantation fluence, the charge state of shallow NV centers estimated from PL and ODMR measurements changes from the NV$^{+}$ to NV$^{-}$ state at an implantation fluence of ${\approx} 1\times10^{12}$ cm$^{-2}$. Moreover, the conductivity decreases with increasing implantation fluence and nearly disappears in a similar implantation fluence range. These results, together with the simulated implantation fluence dependences, indicate that the band bending is limited by a negative surface charge density (surface acceptor density) of ${\approx} 1\times10^{12}$ cm$^{-2}$ under our experimental conditions. This means that band bending does not necessarily proceed until the surface potential energy of diamond reaches the work-function difference between the diamond and surface acceptor material as is generally thought to occur. In addition, our results have also shown spatial inhomogeneities in the surface conductivity and the charge state of the NV centers when the implantation fluence is close to the negative surface charge density.

This work highlights the importance of distinguishing work-function-difference-limited band bending and surface-acceptor-density-limited band bending when treating the surface transfer doping of hydrogen-terminated diamond. This finding will be critical for designing devices based on hydrogen-terminated diamond, particularly for cases in which the density of atmospheric surface acceptors is reduced\cite{Sas22}. Another implication of this work is that controlling the impurity concentration in diamond is important for hydrogen-terminated diamond devices. The impurity concentration has a strong effect on the carrier density, as is indicated by the nitrogen-implantation-fluence dependence of the conductivity. In the case of hydrogen-terminated diamond FETs, the threshold voltage\cite{Oi19} and channel mobility\cite{Sas20} depend on the impurity concentration. In addition, the impurity concentration must be adjusted to prevent reach through and achieve a high breakdown voltage. For this purpose, adjusting the nitrogen concentration by using ion implantation would be a useful approach. Our work thus provides helpful insights that could be used in the development of hydrogen-terminated diamond devices.

\subsection*{Appendix A. NV centers}

NV centers are defects composed of a substitutional nitrogen and an adjacent vacancy in diamond. The charge state of an NV center can be negative (-1), neutral (0), or positive (+1) depending on the Fermi level $E_\mathrm{F}$ at its position. These different charge states can be distinguished by their photonic and magnetic properties.

The NV$^{-}$ state has a spin $S = 1$ and exhibits photoluminescence (PL) with a 637 nm zero phonon line and broad phonon sideband when the diamond is illuminated by a 532-nm excitation light. It is possible to polarize the spin to $m_S = 0$ through illumination of 532-nm light and to make a transition between $m_S = 0$ and $m_S = {\pm}1$ by irradiating the diamond with 2.87 GHz microwaves at room temperature. The spin state can be detected by measuring the PL intensity because the PL intensity for the $m_S = \pm1$ state is weaker than that for $m_S = 0$. These properties of the NV$^{-}$ state allow magnetic resonance to be observed optically (optically detected magnetic resonance; ODMR)\cite{Jel06,Sch14}.

The NV$^{0}$ state shows PL with a 575-nm zero phonon line and broad phonon sideband. The NV$^{0}$ state has a spin $S = 1/2$, but it does not give rise to ODMR. Therefore, the NV$^{-}$ and NV$^{0}$ states can be distinguished by their PL spectra and observation of ODMR specific to NV$^{-}$. The NV$^{+}$ state does not show any PL in the visible wavelength range, which is unlike NV$^{-}$ and NV$^{0}$.

NV centers can be created close to the surface of diamond through nitrogen ion implantation and subsequent annealing. The implantation energy determines the depth distribution of nitrogen and hence, that of the NV centers. This study treats nitrogen and NV centers that are created with an implantation energy of 10 keV and are distributed at depths of around 15 nm. The creation yield of NV centers from implanted nitrogen atoms is ${\approx} 1$\% for an implantation energy of 10 keV\cite{Luh21}.

The calculated NV$^{+}$/NV$^{0}$ and NV$^{0}$/NV$^{-}$ transition levels are at $E_\mathrm{NV}^{+/0} = 1.1$ eV and $E_\mathrm{NV}^{0/-} = 2.7$ eV above the valence band maximum\cite{Dea14}. This means that the NV center is in the NV$^{-}$ state for $E_\mathrm{F} {\textgreater} E_\mathrm{NV}^{0/-}$, in the NV$^{0}$ state for $E_\mathrm{NV}^{+/0} {\textless} E_\mathrm{F} {\textless} E_\mathrm{NV}^{0/-}$, and in the NV$^{+}$ state for $E_\mathrm{F} {\textless} E_\mathrm{NV}^{+/0}$. Therefore, once the charge state of shallow NV centers is determined by PL and ODMR measurements, it can be used to identify the energy range of the Fermi level at the positions of the NV centers. The electric field at the positions of the NV centers can also be estimated from the ODMR frequency.\cite{Dol11,Bro18}

\subsection*{Appendix B. Sample preparation}

Two samples were made from ${\approx} 2.2\times2.2\times0.5$ mm high-purity single-crystal diamond plates with a (001) top surface cut from $4.5\times4.5\times0.5$ mm electronic grade diamond plates (Element Six) with a nitrogen concentration below 5 ppb and boron concentration below 1 ppb. The top surface of the plates was polished by Syntek Co., Ltd. The plates were cleaned in a mixture of sulfuric and nitric acids at $200$\r{}C and rinsed in deionized water. The plate of Sample A was additionally cleaned in organic solvents (acetone and isopropyl alcohol).

The top surface of each diamond plate was divided into four sections (I, II, III, and IV). Nitrogen ion implantation was carried out with different fluences on the four sections by using aluminum foil masks. An implantation with a fluence of $1\times10^{11}$ cm$^{-2}$ was first carried out on all sections of Sample A without a mask. Then, a second implantation with a fluence of $1\times10^{11}$ cm$^{-2}$ was carried out by masking section I. A third implantation with a fluence of $3\times10^{11}$ cm$^{-2}$ was carried out by masking two sections (I and II). Finally, a fourth implantation with a fluence of $5\times10^{11}$ cm$^{-2}$ was carried out by masking three sections (I, II and III). The total fluences for the four sections of Sample A were $1\times10^{11}$, $2\times10^{11}$, $5\times10^{11}$, and $1\times10^{12}$ cm$^{-2}$. Implantations were similarly carried out for Sample B with fluences an order of magnitude larger than those for Sample A. The total fluences for the four sections of Sample B were $1\times10^{12}$, $2\times10^{12}$, $5\times10^{12}$, and $1\times10^{13}$ cm$^{-2}$. After the ion implantation, the plates were cleaned in the acid mixture and rinsed in deionized water as before. The plates were then annealed in vacuum at $1000$\r{}C for two hours to form NV centers. After that, the plates were cleaned in the acid mixture and rinsed in deionized water again.

After formation of NV centers, the diamond plates were exposed to hydrogen plasma in a microwave-plasma-assisted CVD chamber (Seki Technotron, AX5000) to make the surface hydrogen terminated. The hydrogen gas pressure, flow rate, microwave power, treatment time and temperature were $10$ Torr, $50$ sccm, $100$ W, $30$ min, and ${\textless} 600$\r{}C (the detection limit of our pyrometer), respectively. Soft hydrogen plasma was used to avoid reducing the fluorescence of the NV centers\cite{Sta12}. The quality of hydrogen termination made under the above condition was evaluated through contact angle measurements. The contact angle was found to be ${\approx} 90$\r{}, which is characteristics to the hydrophobic nature of the hydrogen-terminated surface\cite{Ma20}. The hydrogen plasma treatment was carried out separately on the two samples. After the treatment, the samples were exposed to air for approximately 2 days before the measurements were started.

\subsection*{Appendix C. Measurement setup}

PL imaging and ODMR measurements were performed using a custom-built confocal microscope system. Details of the setup are described in Ref. \cite{Kag22}. The diamond plate was placed on an epoxy board with a microwave planar ring antenna. A green laser with a wavelength of 532 nm was used for excitation. The laser was focused on the sample thorough an objective lens (Olympus, MPLAPON50X; NA 0.95, WD 0.35 mm). The laser intensity in front of the objective lens was controlled in the range $1 - 200$ $\mu$W by using an ND filter. Fluorescence was detected with an avalanche photo diode through a 648-nm long-pass filter in the PL-mapping and ODMR measurements. Alternatively, it was guided to a spectrometer (Teledyne Princeton Instruments, SpectraPro HRS-300) through a 561-nm long-pass filter. To obtain cw-ODMR spectra, the fluorescence intensity was measured by continuously irradiating the sample with a green laser and sweeping the microwave frequency. In the Rabi-oscillation measurements, pulsed microwaves at the resonance frequency ($m_S = -1$) determined by cw-ODMR were irradiated with the sequence described in Ref. \cite{Kag22}. A magnetic field of ${\approx} 2$ mT was applied to the sample by using a permanent magnet, which led to peak splitting in the ODMR spectra.

The surface conductivity was measured with a two-terminal method. Two prober needles made of Au-based alloy and with a tip radius of 50 $\mu$m were put in contact with the hydrogen-terminated surface in the central region of each section (I, II, III, and IV) of Sample A and B. The distance between the probe contacts was ${\approx} 150$ $\mu$m. Current-voltage characteristics were measured with a source-measure unit (Keysight Technologies, B2902A) for applying a voltage between $0.1$ and $-0.1$ V and a current preamplifier (DL Instruments, 1211) for measuring the current. The current-voltage characteristics were linear, and the conductance was obtained from linear fits.

\vspace{0.6truecm}
\textbf{CRediT authorship contribution statement}

\textbf{Taisuke Kageura:} Investigation, Methodology, Formal Analysis, Writing - review \& editing. \textbf{Yosuke Sasama:} Methodology, Resources. \textbf{Keisuke Yamada:} Resources. \textbf{Kosuke Kimura:} Resources. \textbf{Shinobu Onoda:} Investigation, Methodology, Resources. \textbf{Yamaguchi Takahide:} Conceptualization, Investigation, Methodology, Formal Analysis, Software, Resources, Writing - original draft, Supervision.

\vspace{0.6truecm}
\textbf{Declaration of competing interest}

The authors declare that they have no known competing financial interests or personal relationships that could have appeared to influence the work reported in this paper.

\begin{acknowledgments}

We thank Y. Kubo and J. Isoya for allowing the use of the NV$^{0}$ reference samples. We thank J. Inoue and T. Teraji for their helpful discussions and K. Hino, T. Uchihashi and T. Ando for their support. We also thank M. Monish for correcting the draft. This study was financially supported by JSPS KAKENHI (Grants No. 19H02605, 22H01962, and 23H01429).

\end{acknowledgments}

\begin{figure}
\includegraphics[width=12truecm]{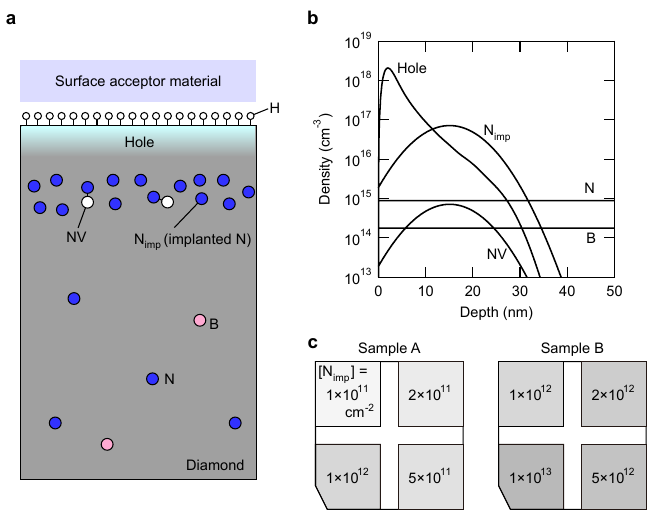}
\caption{\textbf{a} Schematic diagram of hydrogen-terminated diamond with implanted nitrogen and NV centers. \textbf{b} Example depth distributions of implanted nitrogen (N$_\mathrm{imp}$), NV centers (NV), background nitrogen (N) and boron (B) assumed in the band-bending calculation, and calculated depth distribution of holes. The areal density of implanted nitrogen is assumed to be $1\times10^{11}$ cm$^{-2}$. \textbf{c} Schematic diagram of two samples prepared in this work. Sample A has four sections with nitrogen implantation fluences of $1\times10^{11}$, $2\times10^{11}$, $5\times10^{11}$, and $1\times10^{12}$ cm$^{-2}$. Sample B has four sections with nitrogen implantation fluences of $1\times10^{12}$, $2\times10^{12}$, $5\times10^{12}$, and $1\times10^{13}$ cm$^{-2}$. The surfaces of the samples are hydrogen-terminated.}
\end{figure}

\begin{figure}
\includegraphics[width=15truecm]{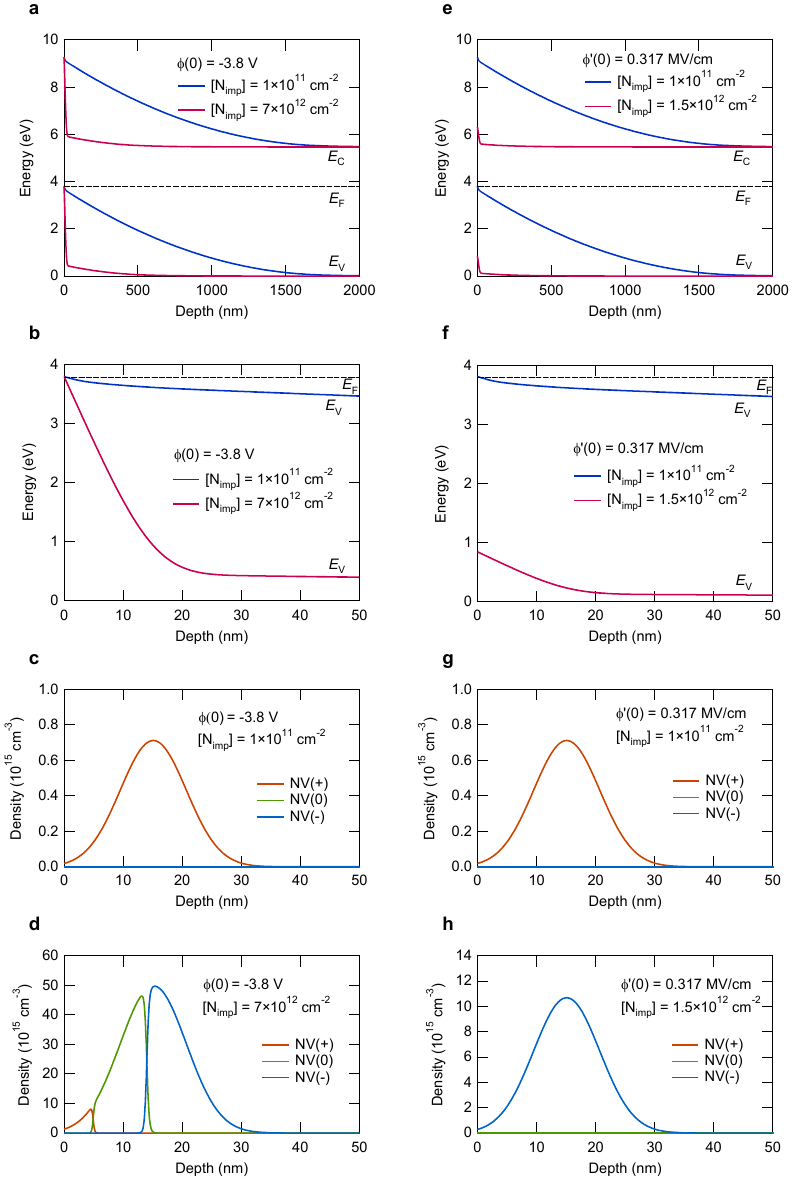}
\end{figure}
\newpage
\begin{figure}
\caption{Calculated results for the constant $\phi(0)$ boundary condition ($\phi(0) = -3.8$ V; \textbf{a}-\textbf{d}) and constant $\phi'(0)$ boundary condition ($\phi'(0) = 0.317$ MV/cm, corresponding to a negative surface charge density of $1\times10^{12}$ cm$^{-2}$; \textbf{e}-\textbf{h}). \textbf{a, b} Energy-band diagrams for nitrogen implantation fluences $\left[\mathrm{N}_\mathrm{imp}\right]$ of $1\times10^{11}$ and $7\times10^{12}$ cm$^{-2}$ in a wide depth range (\textbf{a}) and in a narrow range near the surface (\textbf{b}). \textbf{c, d} Charge state distribution of NV centers for $\left[\mathrm{N}_\mathrm{imp}\right] = 1\times10^{11}$ cm$^{-2}$ (\textbf{c}) and for $\left[\mathrm{N}_\mathrm{imp}\right] = 7\times10^{12}$ cm$^{-2}$ (\textbf{d}). \textbf{e, f} Energy-band diagrams for nitrogen implantation fluences $\left[\mathrm{N}_\mathrm{imp}\right]$ of $1\times10^{11}$ and $1.5\times10^{12}$ cm$^{-2}$ in a wide depth range (\textbf{e}) and in a narrow range near the surface (\textbf{f}). \textbf{g, h} Charge state distribution of NV centers for $\left[\mathrm{N}_\mathrm{imp}\right] = 1\times10^{11}$ cm$^{-2}$ (\textbf{g}) and for $\left[\mathrm{N}_\mathrm{imp}\right] = 1.5\times10^{12}$ cm$^{-2}$ (\textbf{h}). $E_\mathrm{C}$, conduction band minimum; $E_\mathrm{V}$, valence band maximum; and $E_\mathrm{F}$, Fermi level.}
\end{figure}

\begin{figure}
\includegraphics[width=15truecm]{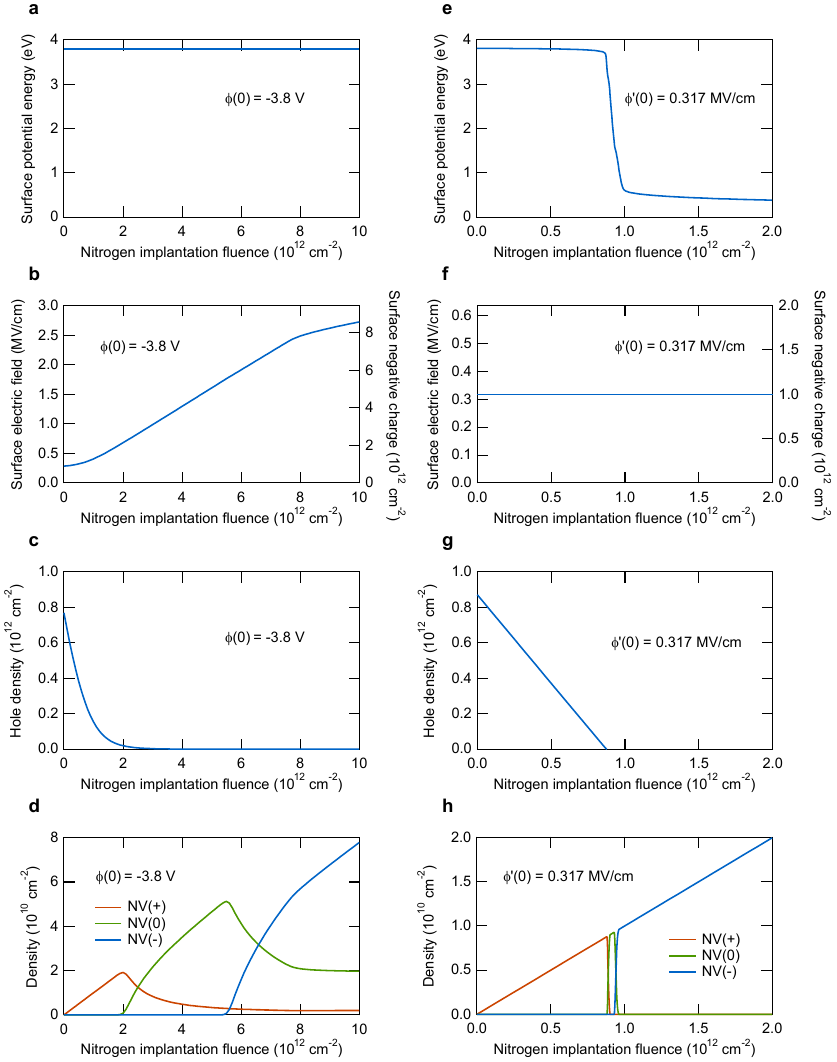}
\caption{Calculated results for the constant $\phi(0)$ boundary condition ($\phi(0) = -3.8$ V; \textbf{a}-\textbf{d}) and constant $\phi'(0)$ boundary condition ($\phi'(0) = 0.317$ MV/cm, corresponding to a negative surface charge density of $1\times10^{12}$ cm$^{-2}$; \textbf{e}-\textbf{h}). The nitrogen-implantation-fluence dependence of the surface potential energy (\textbf{a, e}), surface electric field and negative surface charge density (\textbf{b, f}), hole density (\textbf{c, g}), density of the NV$^{+}$, NV$^{0}$, and NV$^{-}$ states (\textbf{d, h}).}
\end{figure}

\begin{figure}
\includegraphics[width=7.8truecm]{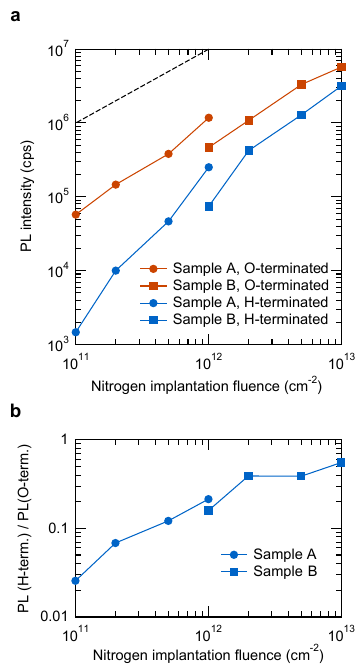}
\caption{\textbf{a} Nitrogen-implantation-fluence dependence of photoluminescence intensity before and after the hydrogen plasma treatment. The diamond surface before the hydrogen plasma treatment was oxygen terminated because of the acid cleaning (Appendix B). The dashed line shows a linear dependence. The intensity for an implantation fluence of ${\le} 1\times10^{12}$ cm$^{-2}$ after hydrogen plasma treatment is multiplied by a factor of 1/10 because the excitation laser intensity is 200 $\mu$W for the implantation fluence, while it is 20 $\mu$W for higher implantation fluences. The intensities for Sample A and B before the hydrogen plasma treatment are multiplied by a factor of 4 and 20 because the excitation laser intensities are 5 $\mu$W and 1 $\mu$W. The reason for the systematic difference in the PL intensity between Sample A and B is unclear, but it may be related to a difference in the quality of the diamond plates. \textbf{b} Ratio of the PL intensity for the hydrogen-terminated surface to that for the oxygen-terminated surface plotted as a function of nitrogen implantation fluence.}
\end{figure}

\begin{figure}
\includegraphics[width=15.5truecm]{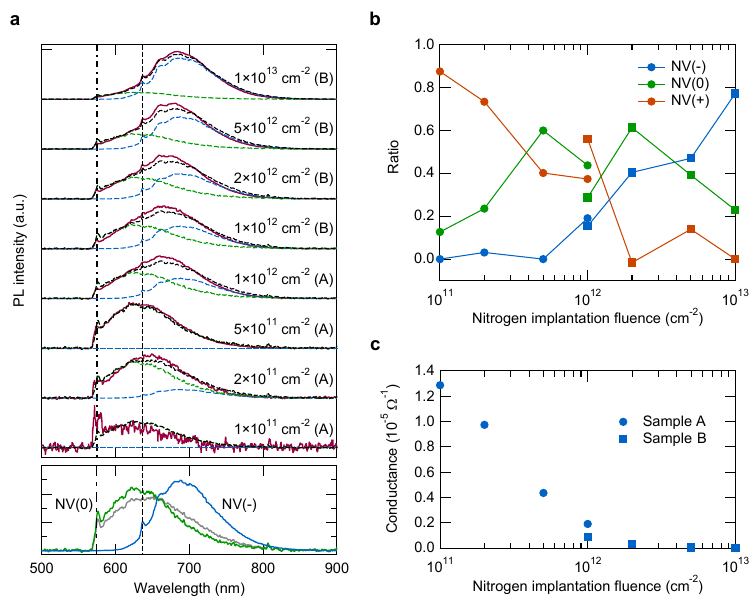}
\caption{\textbf{a} PL spectra (brown curves) from regions with different nitrogen implantation fluences in Sample A and B. Background spectra measured in non-implanted regions (Fig. S4) are subtracted. The vertical dashed dot (dashed) lines indicate the 575-nm (637-nm) zero phonon line of the NV$^{0}$ (NV$^{-}$) centers. The signal level of the spectrum for $\left[\mathrm{N}_\mathrm{imp}\right] = 1\times10^{11}$ cm$^{-2}$ is close to the background (Fig. S4) and the large peaks near 575 nm may be due to imperfect background subtraction. The spectra were fitted with NV$^{0}$ and NV$^{-}$ reference spectra: the green and blue dashed curves are the contributions from NV$^{0}$ and NV$^{-}$ centers and the black dashed curve is the sum of them. The green and blue curves in the bottom panel are the NV$^{0}$ and NV$^{-}$ reference spectra. The gray curve is the spectrum of Sample R$_{\mathrm{NV}(0)}^\mathrm{I}$. (See Section S3 of the Supplementary Material.) \textbf{b} Implantation-fluence dependence of the ratio of the charge states (NV$^{+}$/ NV$^{0}$/ NV$^{-}$) estimated from the PL spectra and intensity. Circles and squares represent Samples A and B. \textbf{c} Implantation-fluence dependence of the conductance.}
\end{figure}

\begin{figure}
\includegraphics[width=15.5truecm]{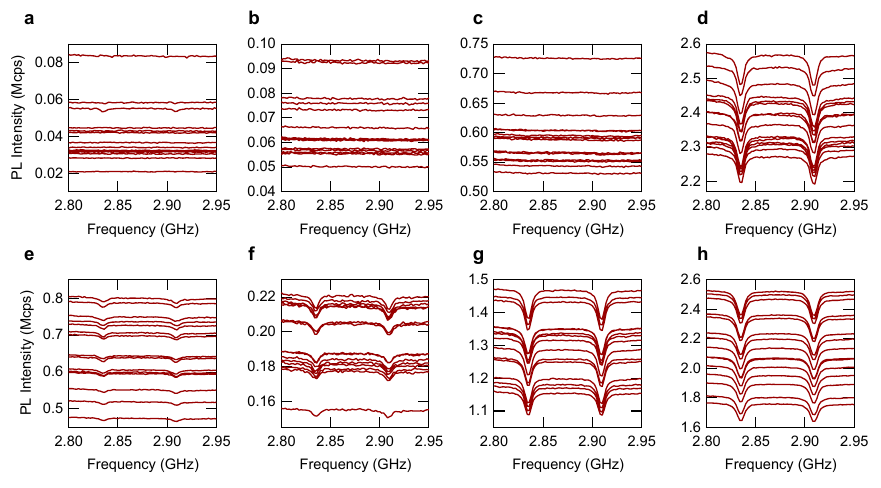}
\caption{\textbf{a-d} ODMR spectra for 15 spots in regions with nitrogen implantation fluences of $1\times10^{11}$ (\textbf{a}), $2\times10^{11}$ (\textbf{b}), $5\times10^{11}$ (\textbf{c}), and $1\times10^{12}$ cm$^{-2}$ (\textbf{d}) in Sample A. \textbf{e-h} ODMR spectra for 15 spots in regions with nitrogen implantation fluences of $1\times10^{12}$ (\textbf{e}), $2\times10^{12}$ (\textbf{f}), $5\times10^{12}$ (\textbf{g}), and $1\times10^{13}$ cm$^{-2}$ (\textbf{h}) in Sample B. The excitation laser intensity is 20 $\mu$W for an implantation fluence of ${\ge} 2\times10^{12}$ cm$^{-2}$ and 200 $\mu$W for lower implantation fluences.}
\end{figure}

\begin{figure}
\includegraphics[width=15.5truecm]{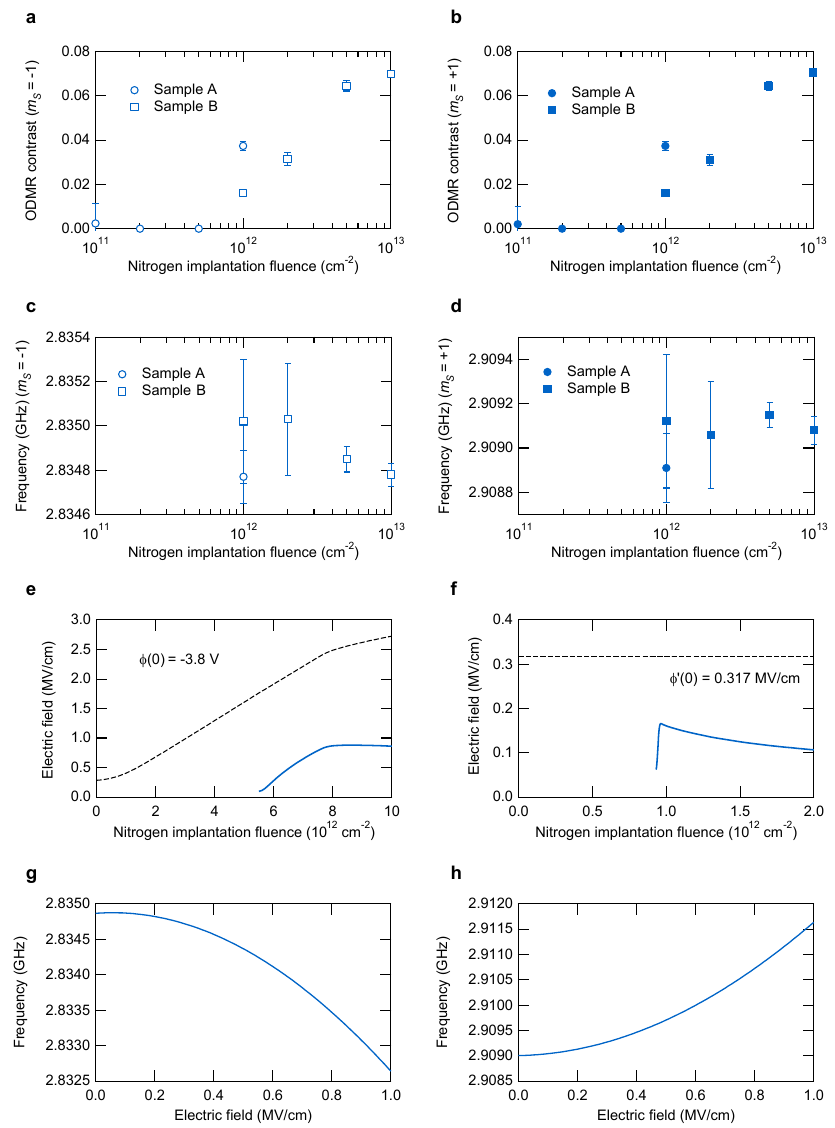}
\end{figure}
\begin{figure}
\caption{\textbf{a,b} ODMR contrasts for $m_S = -1$ (\textbf{a}) and $m_S = +1$ (\textbf{b}) as a function of nitrogen implantation fluence. The contrasts are obtained from Lorentzian fits (Fig. S8) to the ODMR spectra shown in each panel of Fig. 6 and their average over the 15 spectra. The error bars are the standard deviation of the 15 contrasts obtained from the fits. The contrasts for the spectra with no dips (spectra that cannot be fitted to Lorentzian) for fluences ${\le} 5\times10^{11}$ cm$^{-2}$ are set to zero. \textbf{c,d} ODMR frequencies for $m_S = -1$ (\textbf{c}) and $m_S = +1$ (\textbf{d}) as a function of nitrogen implantation fluence. The frequencies and their error bars are obtained from the Lorentzian fits. \textbf{e} Average electric field at the positions of NV$^{-}$ centers (blue solid line) and electric field at the diamond surface (black dashed line) as a function of nitrogen implantation fluence calculated with the constant $\phi(0)$ boundary condition ($\phi(0) = -3.8$ V). \textbf{f} Average electric field at the positions of NV$^{-}$ centers (blue solid line) and electric field at the diamond surface (black dashed line) as a function of nitrogen implantation fluence calculated with the constant $\phi'(0)$ boundary condition ($\phi'(0) = 0.317$ MV/cm, corresponding to a negative surface charge density of $1\times10^{12}$ cm$^{-2}$). \textbf{g,h} Calculated electric-field dependence of the ODMR frequencies $f_{-}$ and $f_{+}$. See Section S5 for the details of the calculation.}
\end{figure}

\clearpage

\vspace{25truecm}

\renewcommand{\theequation}{S\arabic{equation}}
\setcounter{equation}{0}

\textbf{Supplementary Material for ``Surface transfer doping of hydrogen-terminated diamond probed by shallow nitrogen-vacancy centers''}

\subsection*{S1. Calculation of band bending}

We calculated the band bending i.e., the dependence of the potential $\phi(z)$ on the depth $z$ ($z = 0$: diamond surface, $z {\ge} 0$: diamond) for cases in which formation of two-dimensional subbands due to quantum confinement of holes is considered or neglected. When the quantum confinement effect is neglected, only the Poisson equation (\ref{eq:PoissonEq}) shown below is solved. To include the quantum confinement effect, the Poisson and Schr\"odinger equations (\ref{eq:PoissonEq} and \ref{eq:SchrodingerEq}) are solved self-consistently. This paper presents the results of the calculations that include the quantum confinement effect, except for the calculations with the constant $\phi'(0)$ boundary condition for a nitrogen implantation fluence larger than $8.8\times10^{11}$ cm$^{-2}$, where the Fermi level $E_\mathrm{F}$ is far above the valence band maximum and the hole density is very low. The Poisson equation is given by
\begin{eqnarray}
 \frac{d^{2}\phi(z)}{dz^{2}} = -\frac{e}{\epsilon_\mathrm{S}}\left[p(z)+N_{\mathrm{N}^{+}}(z)+N_{\mathrm{N}_{\mathrm{imp}}^{+}}(z)+N_{\mathrm{NV}^{+}}(z)-n(z)-N_{\mathrm{B}^{-}}(z)-N_{\mathrm{NV}^{-}}(z)\right],
\label{eq:PoissonEq}
\end{eqnarray}
where $e$ is the elementary charge and $\epsilon _{s}$ is the static dielectric constant of diamond. Each term on the right-hand side is explained below. The hole and electron densities, $p(z)$ and $n(z)$, are given by\cite{Jen22}
\begin{eqnarray}
 p(z) = 2\left(\frac{2{\pi}k_\mathrm{B}T}{h^2}\right)^{\frac{3}{2}}\left[\left(m^\mathrm{LH}\right)^{\frac{3}{2}}+\left(m^\mathrm{HH}\right)^{\frac{3}{2}}\right]F_{\frac{1}{2}}\left(\frac{E_\mathrm{V}(z)-E_\mathrm{F}}{k_\mathrm{B}T}\right) \nonumber
\end{eqnarray}
\begin{eqnarray}
+2\left(\frac{2{\pi}m^\mathrm{SO}k_\mathrm{B}T}{h^2}\right)^{\frac{3}{2}}F_{\frac{1}{2}}\left(\frac{E_\mathrm{V}(z)-\Delta^\mathrm{SO}-E_\mathrm{F}}{k_\mathrm{B}T}\right),
\label{eq:HoleDensitySemiClassic}
\end{eqnarray}
\begin{eqnarray}
n(z) = 12\left(\frac{2{\pi}k_\mathrm{B}T}{h^2}\right)^{\frac{3}{2}}\left(m_\mathrm{L}m_\mathrm{T}^2\right)^{\frac{1}{2}}F_{\frac{1}{2}}\left(\frac{E_\mathrm{F}-E_\mathrm{C}(z)}{k_\mathrm{B}T}\right),
\end{eqnarray}
\begin{eqnarray}
F_{\frac{1}{2}}(\eta)=\frac{2}{\sqrt{\pi}}\int_{0}^{\infty}du\frac{u^\frac{1}{2}}{1+\exp\left(u-\eta\right)},
\end{eqnarray}
where $k_{B}$ is the Boltzmann constant, $T$ is the absolute temperature, $h$ is the Planck constant, and $F_{\frac{1}{2}}(\eta)$ is the Fermi-Dirac integral. The other parameters and values used in the calculation are shown in Table S1. Note that the valence band maximum $E_\mathrm{V}(z)$ and conduction band minimum $E_\mathrm{C}(z)$ depend on $z$ through the dependence of the potential $\phi(z)$ on $z$.
\begin{eqnarray}
E_{\mathrm{V}}(z) = E_{\mathrm{V}}(z\to\infty) - e\phi(z),
\end{eqnarray}
\begin{eqnarray}
E_{\mathrm{C}}(z) = E_{\mathrm{V}}(z) + E_\mathrm{G},
\end{eqnarray}
where $E_\mathrm{G}$ is the bandgap. Similarly, all the terms on the right-hand side in Eq. \ref{eq:PoissonEq} contain $\phi(z)$. The diamond is assumed to have background nitrogen with a concentration of $N_\mathrm{N} = 5$ ppb ($8.8\times10^{14}$ cm$^{-3}$) and background boron with a concentration of $N_\mathrm{B} = 1$ ppb ($1.76\times10^{14}$ cm$^{-3}$). The concentrations of positively charged background nitrogen and negatively charged background boron are given by
\begin{eqnarray}
N_{\mathrm{N}^{+}}(z) = N_{\mathrm{N}}\frac{1}{1+g_\mathrm{D}\exp\left(\frac{E_\mathrm{F}-E_\mathrm{D}(z)}{k_\mathrm{B}T}\right)},
\end{eqnarray}
\begin{eqnarray}
N_{\mathrm{B}^{-}}(z) = N_{\mathrm{B}}\frac{1}{1+g_\mathrm{A}\exp\left(\frac{E_\mathrm{A}(z)-E_\mathrm{F}}{k_\mathrm{B}T}\right)},
\end{eqnarray}
\begin{eqnarray}
g_\mathrm{D} = 2,
\end{eqnarray}
\begin{eqnarray}
g_\mathrm{A} = 4+2\exp\left(-\frac{\Delta^\mathrm{SO}}{k_\mathrm{B}T}\right),
\end{eqnarray}
\begin{eqnarray}
E_{\mathrm{D}}(z) = E_{\mathrm{C}}(z) - E_\mathrm{D},
\end{eqnarray}
\begin{eqnarray}
E_{\mathrm{A}}(z) = E_{\mathrm{V}}(z) + E_\mathrm{A},
\end{eqnarray}
where $g_\mathrm{D}$ and $g_\mathrm{A}$ are the degeneracy factors\cite{Fon99}, $E_\mathrm{D}$ and $E_\mathrm{A}$ are the ionization energy of donors (nitrogen) and acceptors (boron). The Fermi level $E_\mathrm{F}$, which is independent of $z$, is obtained by solving the charge-neutrality equation for bulk diamond:
\begin{eqnarray}
p(z{\to}{\infty})+N_{\mathrm{N}^{+}}(z{\to}{\infty})-n(z{\to}{\infty})-N_{\mathrm{B}^{-}}(z{\to}{\infty}) = 0.
\end{eqnarray}
The concentration $N_{\mathrm{N}_{\mathrm{imp}}}(z)$ of implanted nitrogen approximately has a Gaussian distribution:
\begin{eqnarray}
N_{\mathrm{N}_{\mathrm{imp}}}(z) = n_{\mathrm{N}_{\mathrm{imp}}}\left(1-R_\mathrm{NV/N}\right)\frac{1}{\sqrt{2\pi}\sigma_\mathrm{imp}}\exp\left(-\frac{\left(z-z_\mathrm{imp}\right)^2}{2{\sigma_\mathrm{imp}}^2}\right),
\end{eqnarray}
where $n_{\mathrm{N}_{\mathrm{imp}}}$ is the nitrogen implantation fluence ($ = \left[\mathrm{N}_\mathrm{imp}\right]$), $R_\mathrm{NV/N}$ is the creation yield of NV centers from implanted nitrogen atoms, $z_\mathrm{imp}$ is the mean depth, and $\sigma_\mathrm{imp}$ is the standard deviation of the distribution. The creation yield $R_\mathrm{NV/N}$ is ${\approx} 1$\% for an implantation energy of 10 keV\cite{Luh21}. We use the values of $z_\mathrm{imp}$ and $\sigma_\mathrm{imp}$ obtained from a Gaussian fit to the nitrogen distribution predicted by a SRIM (Stopping and Range of Ions in Matter) simulation\cite{Zie10} for an implantation energy of 10 keV.
The concentration of ionized implanted nitrogen is given by
\begin{eqnarray}
N_{\mathrm{N}_{\mathrm{imp}}^{+}}(z) = N_{\mathrm{N}_{\mathrm{imp}}}(z)\frac{1}{1+g_\mathrm{D}\exp\left(\frac{E_\mathrm{F}-E_\mathrm{D}(z)}{k_\mathrm{B}T}\right)}.
\end{eqnarray}
The concentration $N_{\mathrm{NV}}(z)$ of the NV center is assumed to have the same distribution as the implanted nitrogen. It is given by
\begin{eqnarray}
N_{\mathrm{NV}}(z) = n_{\mathrm{N}_{\mathrm{imp}}}R_\mathrm{NV/N}\frac{1}{\sqrt{2\pi}\sigma_\mathrm{imp}}\exp\left(-\frac{\left(z-z_\mathrm{imp}\right)^2}{2{\sigma_\mathrm{imp}}^2}\right).
\end{eqnarray}
The concentrations of positively charged, neutral, and negatively charged NV centers are given by
\begin{eqnarray}
N_{\mathrm{NV}^{+}}(z) = N_{\mathrm{NV}}(z)\frac{1}{1+4\exp\left(\frac{E_\mathrm{F}-E_\mathrm{NV}^{+/0}(z)}{k_\mathrm{B}T}\right)+2\exp\left(\frac{2E_\mathrm{F}-E_\mathrm{NV}^{+/0}(z)-E_\mathrm{NV}^{0/-}(z)}{k_\mathrm{B}T}\right)},
\label{eq:NposNV}
\end{eqnarray}
\begin{eqnarray}
N_{\mathrm{NV}^{0}}(z) = N_{\mathrm{NV}}(z)\frac{4\exp\left(\frac{E_\mathrm{F}-E_\mathrm{NV}^{+/0}(z)}{k_\mathrm{B}T}\right)}{1+4\exp\left(\frac{E_\mathrm{F}-E_\mathrm{NV}^{+/0}(z)}{k_\mathrm{B}T}\right)+2\exp\left(\frac{2E_\mathrm{F}-E_\mathrm{NV}^{+/0}(z)-E_\mathrm{NV}^{0/-}(z)}{k_\mathrm{B}T}\right)},
\label{eq:NneuNV}
\end{eqnarray}
\begin{eqnarray}
N_{\mathrm{NV}^{-}}(z) = N_{\mathrm{NV}}(z)\frac{2\exp\left(\frac{2E_\mathrm{F}-E_\mathrm{NV}^{+/0}(z)-E_\mathrm{NV}^{0/-}(z)}{k_\mathrm{B}T}\right)}{1+4\exp\left(\frac{E_\mathrm{F}-E_\mathrm{NV}^{+/0}(z)}{k_\mathrm{B}T}\right)+2\exp\left(\frac{2E_\mathrm{F}-E_\mathrm{NV}^{+/0}(z)-E_\mathrm{NV}^{0/-}(z)}{k_\mathrm{B}T}\right)},
\label{eq:NnegNV}
\end{eqnarray}
\begin{eqnarray}
E_\mathrm{NV}^{+/0}(z) = E_{\mathrm{V}}(z) + E_\mathrm{NV}^{+/0},
\end{eqnarray}
\begin{eqnarray}
E_\mathrm{NV}^{0/-}(z) = E_{\mathrm{V}}(z) + E_\mathrm{NV}^{0/-},
\end{eqnarray}
where $E_\mathrm{NV}^{+/0}$ and $E_\mathrm{NV}^{0/-}$ are the transition energy levels of NV$^{+}$/NV$^{0}$ and NV$^{0}$/NV$^{-}$. $E_\mathrm{NV}^{+/0}$ and $E_\mathrm{NV}^{0/-}$ are calculated to be 1.1 and 2.7 eV above the valence band maximum\cite{Dea14}. The degeneracy factor of 4 of the NV$^{0}$ state (in the denominator of the righthand side in Eqs. \ref{eq:NposNV}, \ref{eq:NneuNV}, and \ref{eq:NnegNV}) comes from the spin and orbital ($e_x$ and $e_y$) degrees of freedom. The degeneracy factor of 2 of the NV$^{-}$ state comes from the orbital ($e_x$ and $e_y$) degrees of freedom.

The quantum confinement effect of holes is treated by solving the Poisson equation (Eq. \ref{eq:PoissonEq}) and Schr\"odinger equation,
\begin{eqnarray}
 \left[-\frac{\hbar^{2}}{2m_{z}^{i}}\frac{d^{2}}{dz^{2}} + e\phi(z) (+\Delta^\mathrm{SO}) - E_{n}^{i}\right]\Psi_{n}^{i}(z) = 0,
\label{eq:SchrodingerEq}
\end{eqnarray}
in a self-consistent manner. ($\Delta^\mathrm{SO}$ is taken into account only in the calculation for split-off holes.) In this case, the hole density $p(z)$ is given by
\begin{eqnarray}
 p(z) = \sum_{i,n}p_{n}^{i}\left|\Psi_{n}^{i}(z)\right|^{2},
\end{eqnarray}
\begin{eqnarray}
 p_{n}^{i} = \frac{m_{//}^{i}k_{B}T}{\pi\hbar^{2}}\ln\left[1+\exp\left(\frac{E_{F}-E_{n}^{i}}{k_{B}T}\right)\right],
\end{eqnarray}
instead of Eq. \ref{eq:HoleDensitySemiClassic}.

\subsection*{S2. Applicable boundary condition in our experiments}

In our experiments, the applicable boundary condition at the diamond surface may depend on the nitrogen implantation fluence. For example, it is possible that the constant $\phi(0)$ boundary condition is applicable to a low implantation fluence where not all of the surface acceptors are ionized, whereas the constant $\phi'(0)$ boundary condition is applicable to a high implantation fluence where all of the surface acceptors are ionized. The calculation under the constant $\phi'(0)$ boundary condition with a constant negative surface charge density of $1\times10^{12}$ cm$^{-2}$ leads to the result that $-e\phi(0)$ is larger than 3.8 eV for implantation fluences less than $4.7\times10^{11}$ cm$^{-2}$. In this implantation fluence range, the calculation under the constant $\phi(0)$ boundary condition ($-e\phi(0) = 3.8$ eV) provides a negative surface charge density of less than $1\times10^{12}$ cm$^{-2}$. Therefore, it is possible that $\phi(0)$ is constant ($= -3.8$ V) for $\left[\mathrm{N}_\mathrm{imp}\right] {\textless} 4.7\times10^{11}$ cm$^{-2}$ and $\phi'(0)$ is constant ($n_\mathrm{SA}^{-} = 1\times10^{12}$ cm$^{-2}$) for $\left[\mathrm{N}_\mathrm{imp}\right] {\textgreater} 4.7\times10^{11}$ cm$^{-2}$. As shown in the main text, our experimental results are in better agreement with the calculation under the constant $\phi'(0)$ boundary condition than they are with the calculation under the constant $\phi(0)$ boundary condition. However, the constant $\phi(0)$ boundary condition may be appropriate for low implantation fluences ${\le} 2\times10^{11}$ cm$^{-2}$.

\subsection*{S3. NV$^{0}$ and NV$^{-}$ reference spectra}

The reference samples for the NV$^{0}$ and NV$^{-}$ spectra were prepared by controlling the density of donors (nitrogen; P1 center) and acceptors (boron). A NV$^{-}$ reference sample (Sample R$_{\mathrm{NV}(-)}$) was made from a Ib-type HPHT (001) single-crystalline diamond plate (Sumitomo Electric Industries) with a nitrogen concentration of approximately a hundred ppm. The diamond plate was irradiated with an electron beam with a fluence of $10^{18}$ cm$^{-2}$ and energy of $2$ MeV for creating vacancies and annealed in vacuum at $1000$\r{}C for $2$ h for forming NV centers. The P1 concentration evaluated by ESR measurements after electron beam irradiation and before annealing was 134 ppm. The [NV]/[P1] ratio after annealing for the above condition was estimated to be $1 - 2$\%. The P1 concentration is sufficiently large to make most NV centers be in the negative charge state\cite{Ishi22}. The PL spectrum for the NV$^{-}$ reference sample was measured with a low excitation power to avoid a photo-induced charge state conversion. NV$^{0}$ reference samples (Samples R$_{\mathrm{NV}(0)}^\mathrm{I}$ and R$_{\mathrm{NV}(0)}^\mathrm{II}$) were prepared using two different methods. Sample R$_{\mathrm{NV}(0)}^\mathrm{I}$ was made from a IIa-type (001) single-crystalline diamond plate (Element Six) with a relatively low nitrogen concentration of 40 ppb (measured by ESR). The isolated nitrogen density was decreased further by converting nitrogen into a NV center; the diamond plate was irradiated with an electron beam with a fluence of $8.5\times10^{17}$ cm$^{-2}$ and energy of 2 MeV and annealed in vacuum at $1000$\r{}C for $2$ h. The P1 concentration after the irradiation and annealing was below the detection limit of the ESR equipment. The low P1 concentration favors the NV$^{0}$ state\cite{Shi21}. Sample R$_{\mathrm{NV}(0)}^\mathrm{II}$ was made from a HPHT boron-doped diamond plate with a resistivity of $10^{2}-10^{3}$ $\Omega$cm. The diamond was irradiated with an electron beam with a fluence of $10^{18}$ cm$^{-2}$ and energy of $2$ MeV and annealed in vacuum at $1000$\r{}C for 2 h.

The normalized PL spectra of the NV$^{0}$ and NV$^{-}$ reference samples are shown in Fig. S5. In a short wavelength range ${\textless} 600$ nm, the PL intensity of the NV$^{-}$ reference sample (Sample R$_{\mathrm{NV}(-)}$) is low and indicates that the contribution of NV$^{0}$ PL is smaller than ${\approx} 1$\%. Therefore, we chose the spectrum of Sample R$_{\mathrm{NV}(-)}$ as the NV$^{-}$ reference spectrum. The PL spectra of the NV$^{0}$ reference samples prepared with the different methods were in good agreement with each other. We selected Sample R$_{\mathrm{NV}(0)}^\mathrm{I}$, the spectrum of which has a clearer zero phonon line, as the NV$^{0}$ reference sample. The PL intensities of the NV$^{0}$ reference sample in a wavelength range of $700 - 800$ nm are nearly the same as 0.4 times the intensities of the NV$^{-}$ reference spectrum. Therefore, one cannot rule out the possibility that the spectrum of Sample R$_{\mathrm{NV}(0)}^\mathrm{I}$ contains PL from NV$^{-}$ centers. We supposed three cases in which 0\%, 20\% and 40\% of the PL of Sample R$_{\mathrm{NV}(0)}^\mathrm{I}$ comes from NV$^{-}$ centers. We subtracted the NV$^{-}$ contribution from the spectrum of Sample R$_{\mathrm{NV}(0)}^\mathrm{I}$ and used it as the NV$^{0}$ reference spectrum for fitting. (After the subtraction of the NV$^{-}$ contribution, the spectrum is normalized so that the integrated intensity is unchanged.) Figures S6a-c show the fitting results for the three cases. Better fits (for $\left[\mathrm{N}_\mathrm{imp}\right] = 1\times10^{11}$ and $5\times10^{11}$ cm$^{-2}$) are obtained for the ${\ge}20$\% NV$^{-}$ contribution to the spectrum of Sample R$_{\mathrm{NV}(0)}^\mathrm{I}$. It is also worth noting that the NV$^{0}$ reference spectrum for the 20\% case is in the best agreement with the spectrum of a single NV$^{0}$ reported in Ref. \cite{Ron10}. Figure S6d shows the relative contributions of NV$^{0}$ and NV$^{-}$ to the PL spectra obtained by the fitting. Figure S6e shows the NVH/NV$^{+}$/NV$^{0}$/NV$^{-}$ relative populations calculated with the data shown in Figs. 4b and S6d. (See Section S4 for the calculation.) Figures S6f shows the estimate of NV$^{+}$/NV$^{0}$/NV$^{-}$ relative populations. Figures S6d-f show the range of uncertainty caused by the uncertainty of the NV$^{-}$ contribution to the PL spectrum of Sample R$_{\mathrm{NV}(0)}^\mathrm{I}$. Figures 5a and 5b in the main text shows the results for the case in which the NV$^{-}$ contribution to the spectrum of Sample R$_{\mathrm{NV}(0)}^\mathrm{I}$ is assumed to be 20\%.

\subsection*{S4. Analysis of charge states with photoluminescence}

We evaluated the relative population of the charge states (NV$^{+}$/NV$^{0}$/NV$^{-}$) below the hydrogen-terminated surface from the observed PL intensities and spectra by using the following equations. The relative populations of NV$^{+}$, NV$^{0}$, and NV$^{-}$ below the oxygen-terminated surface (before hydrogenation), $p_{+}^\mathrm{O}$, $p_{0}^\mathrm{O}$, and $p_{-}^\mathrm{O}$, follow
\begin{eqnarray}
p_{+}^\mathrm{O} + p_{0}^\mathrm{O} + p_{-}^\mathrm{O} = 1.
\label{eq:OtermRatio}
\end{eqnarray}
We assume that some of the NV centers are transformed to NVH after hydrogenation. The relative populations of NVH, NV$^{+}$, NV$^{0}$, and NV$^{-}$ below the hydrogen-terminated surface, $p_\mathrm{NVH}^\mathrm{H}$, $p_{+}^\mathrm{H}$, $p_{0}^\mathrm{H}$, and $p_{-}^\mathrm{H}$, follow
\begin{eqnarray}
p_\mathrm{NVH}^\mathrm{H} + p_{+}^\mathrm{H} + p_{0}^\mathrm{H} + p_{-}^\mathrm{H} = 1.
\label{eq:HtermRatio} 
\end{eqnarray}
The ratio of the PL intensity for the hydrogen-terminated surface to that for the oxygen-terminated surface, $I_\mathrm{PL}^\mathrm{H}/I_\mathrm{PL}^\mathrm{O}$ (Fig. 4b), is given by
\begin{eqnarray}
\frac{I_\mathrm{PL}^\mathrm{H}}{I_\mathrm{PL}^\mathrm{O}} = \frac{t_{0}p_{0}^\mathrm{H}{\Gamma}_{0}+t_{-}p_{-}^\mathrm{H}{\Gamma}_{-}}{t_{0}p_{0}^\mathrm{O}{\Gamma}_{0}+t_{-}p_{-}^\mathrm{O}{\Gamma}_{-}},
\label{eq:PLIntensity}
\end{eqnarray}
where $\Gamma_{0}$ and $\Gamma_{-}$ are the photon emission rate of the NV$^{0}$ and NV$^{-}$ centers, $t_{0}$ and $t_{-}$ are the transmittances of the PL signals from NV$^{0}$ and NV$^{-}$ centers through the 648-nm long-pass filter (used in the PL-intensity measurements). The ratio of the NV$^{0}$ and NV$^{-}$ contributions to the PL spectra measured for the hydrogen-terminated surface, $c_{-}^\mathrm{H}/c_{0}^\mathrm{H}$ (Fig. S6d), is given by
\begin{eqnarray}
\frac{c_{-}^\mathrm{H}}{c_{0}^\mathrm{H}} = \frac{p_{-}^\mathrm{H}{\Gamma}_{-}}{p_{0}^\mathrm{H}{\Gamma}_{0}}.
\label{eq:PLSpectra}
\end{eqnarray}

From Eqs. \ref{eq:PLIntensity} and \ref{eq:PLSpectra}, $p_{0}^\mathrm{H}$ and $p_{-}^\mathrm{H}$ can be obtained:
\begin{eqnarray}
p_{0}^\mathrm{H} = \frac{I_\mathrm{PL}^\mathrm{H}}{I_\mathrm{PL}^\mathrm{O}} (t_{0}p_{0}^\mathrm{O}{\Gamma}_{0}+t_{-}p_{-}^\mathrm{O}{\Gamma}_{-})\frac{1}{{\Gamma}_{0}}\frac{c_{0}^\mathrm{H}}{t_{0}c_{0}^\mathrm{H}+t_{-} c_{-}^\mathrm{H}},
\label{eq:p0H}
\end{eqnarray}
\begin{eqnarray}
p_{-}^\mathrm{H} = \frac{I_\mathrm{PL}^\mathrm{H}}{I_\mathrm{PL}^\mathrm{O}} (t_{0}p_{0}^\mathrm{O}{\Gamma}_{0}+t_{-}p_{-}^\mathrm{O}{\Gamma}_{-})\frac{1}{{\Gamma}_{-}}\frac{c_{-}^\mathrm{H}}{t_{0}c_{0}^\mathrm{H}+t_{-} c_{-}^\mathrm{H}}.
\label{eq:pmH}
\end{eqnarray}
By assuming that $p_{+}^\mathrm{H} = 0$ for a nitrogen implantation fluence of $10^{13}$ cm$^{-2}$ and $p_\mathrm{NVH}^\mathrm{H}$ is independent of nitrogen implantation fluence, one can also obtain $p_\mathrm{NVH}^\mathrm{H}$ and $p_{+}^\mathrm{H}$ from Eq. \ref{eq:HtermRatio}:
\begin{eqnarray}
p_\mathrm{NVH}^\mathrm{H} = 1 - p_{0}^\mathrm{H}(\left[\mathrm{N}_\mathrm{imp}\right] = 1\times10^{13} \mathrm{cm}^{-2}) - p_{-}^\mathrm{H}(\left[\mathrm{N}_\mathrm{imp}\right] = 1\times10^{13} \mathrm{cm}^{-2}),
\label{eq:pNVHH}
\end{eqnarray}
\begin{eqnarray}
p_{+}^\mathrm{H} = 1 - p_{0}^\mathrm{H} - p_{-}^\mathrm{H} - p_\mathrm{NVH}^\mathrm{H}.
\label{eq:ppH}
\end{eqnarray}
$p_{0}^\mathrm{H}$ and $p_{-}^\mathrm{H}$ were calculated with Eqs. \ref{eq:p0H} and \ref{eq:pmH} using the data shown in Figs. 4b and S6d. Here, we assumed that $1/\Gamma_{-} = 12$ ns and $1/\Gamma_{0} = 19$ ns for high excitation power conditions\cite{Luo22,Rob19}. We also assumed that $p_{+}^\mathrm{O} = 0$, $p_{0}^\mathrm{O} = 0.2$, $p_{-}^\mathrm{O} = 0.8$ by taking account of photo-induced ionization\cite{Asl13}. The values of $p_{+}^\mathrm{O}$, $p_{0}^\mathrm{O}$, and $p_{-}^\mathrm{O}$ affect the estimate of $p_\mathrm{NVH}^\mathrm{H}$, but do not affect the final result (Figs. 5b and S6f) on the NV$^{+}$/NV$^{0}$/NV$^{-}$ relative populations. The transmittances for the 648-nm long-pass filter were estimated from the NV$^{0}$ and NV$^{-}$ reference spectra: $t_{-} = 0.92$ and $t_{0} = 0.55$, $0.45$, and $0.30$ for the 0\%, 20\%, and 40\% cases described in Section S3. $p_\mathrm{NVH}^\mathrm{H}$ and $p_{0}^\mathrm{H}$ were also obtained with Eqs. \ref{eq:pNVHH} and \ref{eq:ppH}. The obtained $p_\mathrm{NVH}^\mathrm{H}$, $p_{+}^\mathrm{H}$, $p_{0}^\mathrm{H}$, and $p_{-}^\mathrm{H}$ are shown in Fig. S6e. Figures 5b and S6f show the relative populations excluding the NVH defects: $p_{+}^\mathrm{H}/(p_{+}^\mathrm{H}+p_{0}^\mathrm{H}+p_{-}^\mathrm{H})$, $p_{0}^\mathrm{H}/(p_{+}^\mathrm{H}+p_{0}^\mathrm{H}+p_{-}^\mathrm{H})$, and $p_{-}^\mathrm{H}/(p_{+}^\mathrm{H}+p_{0}^\mathrm{H}+p_{-}^\mathrm{H})$.

\subsection*{S5. Calculation of the electric-field-induced shift of the ODMR frequency}

We calculated the electric-field-induced shift of the ODMR frequency by following Ref. \cite{Doh14}. The spin Hamiltonian of the NV center in the presence of magnetic and electric fields ($\vec{B}$, $\vec{E}$) is given by
\begin{eqnarray}
H = (D + k_{z}E_{z})({S_{z}}^2 - 2/3) + {\gamma_{e}}\vec{S}\cdot\vec{B} - k_xE_x({S_x}^2-{S_y}^2) + k_yE_y(S_xS_y + S_yS_x),
\end{eqnarray}
where $\vec{S}$ are the $S = 1$ dimensionless electron spin operators, $D {\approx} 2.87$ GHz is the zero-field splitting, $\gamma_{e} = {g_e}{\mu_\mathrm{B}}/h$ is the electron gyromagnetic ratio, $\mu_\mathrm{B}$ is the Bohr magneton, $g_e {\approx} 2.003$ is the electron g-factor, $h$ is the Planck constant, $\vec{B}$ and $\vec{E}$ are magnetic and electric fields, and $k_z = 3.5$ kHz $\mu$m V$^{-1}$ and $k_x = k_y = k_\perp = 170$ kHz $\mu$m V$^{-1}$ are the electric susceptibility parameters. $xyz$ is the coordinate frame defined in Fig. 2 of Ref \cite{Doh14}, where $z$ is the direction joining nitrogen and vacancy and $x$ is orthogonal to $z$ and contained within one of the center's three reflection planes. The ODMR frequencies $f_{\pm}$ are approximately
\begin{eqnarray}
f_{\pm} = D + k_{z}E_{z} + 3\Lambda \pm \sqrt{R^2-{\Lambda} R \sin{\alpha} \cos{\beta} + \Lambda^2},
\end{eqnarray}
where $\Lambda = {\gamma_{e}}^2{B_\perp}^2/2D$, $R=\sqrt{{\gamma_{e}}^2R^2+{k_\perp}^2{E_\perp}^2}$, $B_\perp = \sqrt{{B_x}^2+{B_y}^2}$, $E_\perp = \sqrt{{E_x}^2+{E_y}^2}$, $\tan{\alpha} = {k_\perp}{E_\perp}/{\gamma_{e}}B_z$, $\beta = 2\phi_B+\phi_E$, $\tan{\phi_B} = B_y/B_x$, $\tan{\phi_E} = E_y/E_x$. The above expression for $f_{\pm}$ is valid under the conditions, $\Lambda, R << D$. We assume that the magnetic and electric fields are parallel to the [001] crystal direction. Therefore, $\vec{B} = (\sqrt{2/3}|\vec{B}|, 0, |\vec{B}|/\sqrt{3})$ and $\vec{E} = (\sqrt{2/3}|\vec{E}|, 0, |\vec{E}|/\sqrt{3})$. The calculated ODMR frequencies are plotted as a function of electric field in Figs. 7g and 7h. Here, $D = 2.8705$ GHz and $|\vec{B}| = 2.29$ mT were assumed for explaining the observed splitting of the ODMR frequencies.

\newpage

\begin{table}
\begin{tabular}{|c|c|c|}
\hline
Temperature & $T$ & 300 K \\
\hline
Bulk donor (nitrogen) density & $N_\mathrm{N}$ & $8.8\times10^{14}$ cm$^{-3}$ \\
\hline
Bulk acceptor (boron) density & $N_\mathrm{B}$ & $1.76\times10^{14}$ cm$^{-3}$ \\
\hline
Donor (nitrogen) ionization energy & $E_\mathrm{D}$ & $1.7$ eV \\
\hline
Acceptor (boron) ionization energy & $E_\mathrm{A}$ & $0.37$ eV \\
\hline
NV$^{+}$/NV$^{0}$ transition level (relative to VBM) & $E_\mathrm{NV}^{+/0}$ & $1.1$ eV \\
\hline
NV$^{0}$/NV$^{-}$ transition level (relative to VBM) & $E_\mathrm{NV}^{0/-}$ & $2.7$ eV \\
\hline
Mean depth of implanted nitrogen & $z_\mathrm{imp}$ & 15.1 nm \\
\hline
Standard deviation of implanted nitrogen distribution & $\sigma_\mathrm{imp}$ & $5.6$ nm \\
\hline
Creation yield of NV centers from implanted N atoms & $R_\mathrm{NV/N}$ & $0.01$ \\
\hline
Bandgap & $E_\mathrm{G}$ & 5.47 eV \\
\hline
Dielectric constant & $\epsilon_\mathrm{S}/\epsilon_{0}$ & $5.7$ \\
\hline
Spin-orbit splitting energy & $\Delta^\mathrm{SO}$ & $6$ meV \\
\hline
Effective mass of heavy hole & $m^\mathrm{HH}/m_{0}$ & $0.67$ \\
\hline
Effective mass of light hole & $m^\mathrm{LH}/m_{0}$ & 0.26 \\
\hline
Effective mass of split-off hole & $m^\mathrm{SO}/m_{0}$ & 0.375 \\
\hline
Effective mass of heavy hole along [001] & $m_{z}^\mathrm{HH}/m_{0}$ & 0.441 \\
\hline
Effective mass of light hole along [001] & $m_{z}^\mathrm{LH}/m_{0}$ & 0.325 \\
\hline
Effective mass of split-off hole along [001] & $m_{z}^\mathrm{SO}/m_{0}$ & 0.375 \\
\hline
Effective mass of heavy hole in the (001) plane & $m_{//}^\mathrm{HH}/m_{0}$ & 0.288 \\
\hline
Effective mass of light hole in the (001) plane & $m_{//}^\mathrm{LH}/m_{0}$ & 0.536 \\
\hline
Effective mass of split-off hole in the (001) plane & $m_{//}^\mathrm{SO}/m_{0}$ & 0.375 \\
\hline
Longitudinal effective mass of electron & $m_\mathrm{L}/m_{0}$ & 1.56 \\
\hline
Transverse effective mass of electron & $m_\mathrm{T}/m_{0}$ & 0.28 \\
\hline
\end{tabular}
\begin{flushleft}
{\footnotesize Table S1: Parameters used in the calculation. $\epsilon _{0}$ is the vacuum permittivity. $m_{0}$ is the rest mass. The effective masses are from the experimental results of Ref. \cite{Nak13}.}
\end{flushleft}
\label{table:parameters}
\end{table}

\newpage

\begin{figure}
\includegraphics[width=15truecm]{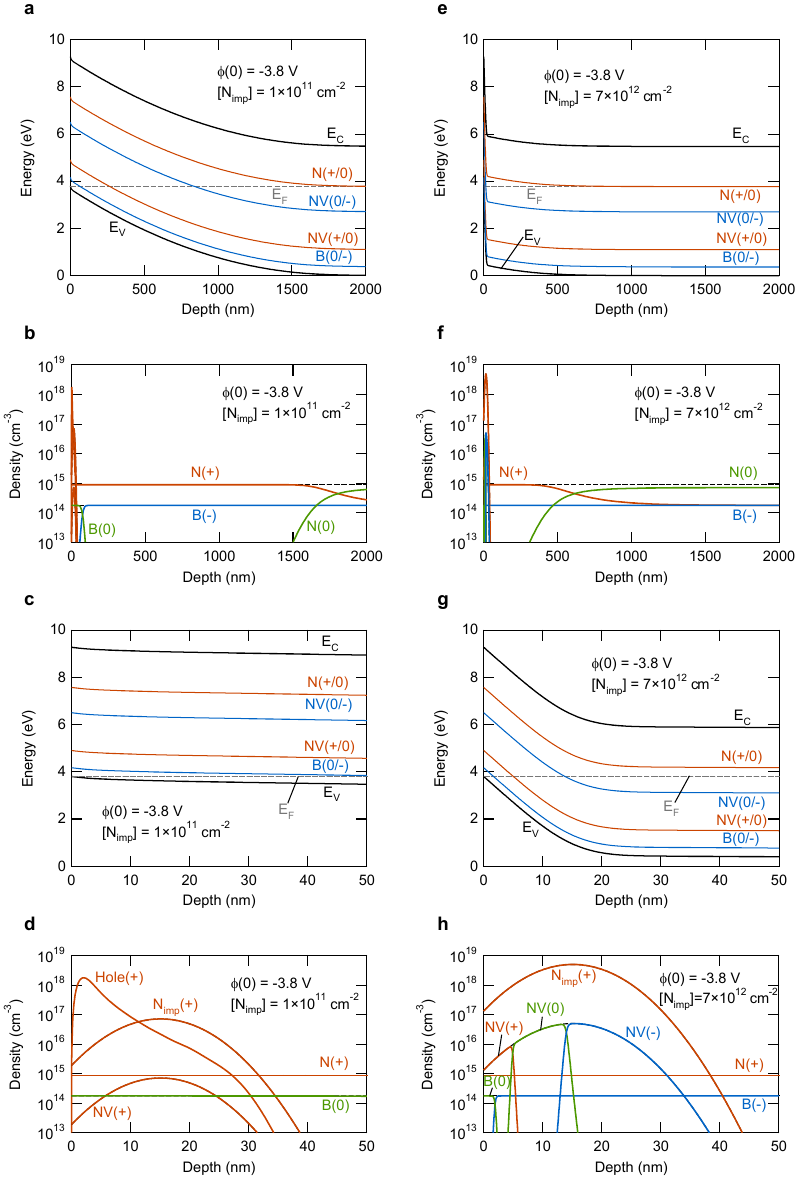}
\end{figure}
\newpage
\begin{figure}
\begin{flushleft}
{\footnotesize Figure S1: Band diagram, charge transition levels and charge state distribution of nitrogen, boron, and NV centers calculated with the constant $\phi(0)$ boundary condition ($\phi(0) = -3.8$ V). \textbf{a-d} Band diagrams and charge transition levels (\textbf{a,c}) and charge state distribution (\textbf{b,d}) for a nitrogen implantation fluence $\left[\mathrm{N}_\mathrm{imp}\right]$ of $1\times10^{11}$ cm$^{-2}$ in a wide depth range (\textbf{a,b}) and in a narrow range near the surface (\textbf{c,d}). \textbf{e-h} Band diagrams and charge transition levels (\textbf{e,g}) and charge state distribution (\textbf{f,h}) for a nitrogen implantation fluence $\left[\mathrm{N}_\mathrm{imp}\right]$ of $7\times10^{12}$ cm$^{-2}$ in a wide depth range (\textbf{e,f}) and in a narrow range near the surface (\textbf{g,h}). $E_\mathrm{C}$, conduction band minimum; $E_\mathrm{V}$, valence band maximum; and $E_\mathrm{F}$, Fermi level.}
\end{flushleft}
\end{figure}

\begin{figure}
\includegraphics[width=15truecm]{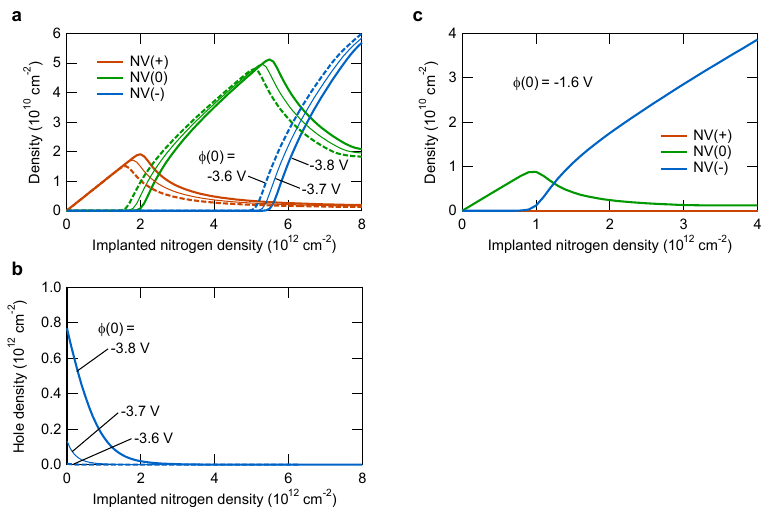}
\begin{flushleft}
{\footnotesize Figure S2 Calculated results for the constant $\phi(0)$ boundary condition for different values of $\phi(0)$. \textbf{a} NV$^{+}$, NV$^{0}$, and NV$^{-}$ density as a function of nitrogen implantation fluence for $\phi(0) = -3.6$, $-3.7$, and $-3.8$ V. \textbf{b} Hole density as a function of nitrogen implantation fluence for $\phi(0) = -3.6$, $-3.7$, and $-3.8$ V. \textbf{c} NV$^{+}$, NV$^{0}$, and NV$^{-}$ density as a function of nitrogen implantation fluence for $\phi(0) = -1.6$ V.}
\end{flushleft}
\end{figure}

\begin{figure}
\includegraphics[width=15truecm]{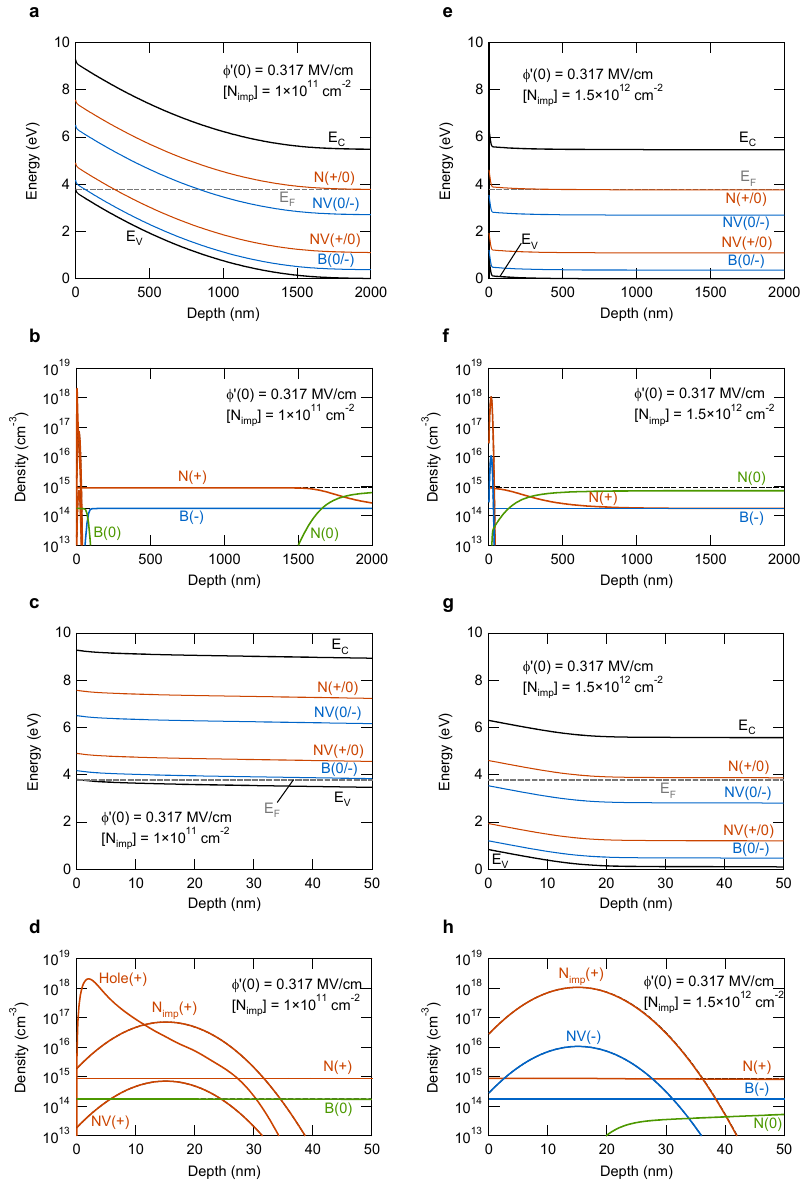}
\end{figure}
\newpage
\begin{figure}
\begin{flushleft}
{\footnotesize Figure S3: Band diagram, charge transition levels and charge state distribution of nitrogen, boron, and NV centers calculated with the constant $\phi'(0)$ boundary condition ($\phi'(0) = 0.317$ MV/cm, corresponding to a negative surface charge density of $1\times10^{12}$ cm$^{-2}$). \textbf{a-d} Band diagrams and charge transition levels (\textbf{a,c}) and charge state distribution (\textbf{b,d}) for a nitrogen implantation fluence $\left[\mathrm{N}_\mathrm{imp}\right]$ of $1\times10^{11}$ cm$^{-2}$ in a wide depth range (\textbf{a,b}) and in a narrow range near the surface (\textbf{c,d}). \textbf{e-h} Band diagrams and charge transition levels (\textbf{e,g}) and charge state distribution (\textbf{f,h}) for a nitrogen implantation fluence $\left[\mathrm{N}_\mathrm{imp}\right]$ of $1.5\times10^{12}$ cm$^{-2}$ in a wide depth range (\textbf{e,f}) and in a narrow range near the surface (\textbf{g,h}). $E_\mathrm{C}$, conduction band minimum; $E_\mathrm{V}$, valence band maximum; and $E_\mathrm{F}$, Fermi level.}
\end{flushleft}
\end{figure}

\begin{figure}
\includegraphics[width=6truecm]{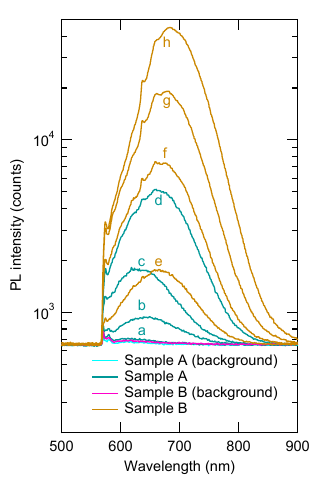}
\end{figure}
\begin{figure}
\begin{flushleft}
{\footnotesize Figure S4 \textbf{a} Raw data of PL spectra measured in regions with nitrogen implantation fluences of $1\times10^{11}$ (a), $2\times10^{11}$ (b), $5\times10^{11}$ (c), and $1\times10^{12}$ cm$^{-2}$ (d) in Sample A and $1\times10^{12}$ (e), $2\times10^{12}$ (f), $5\times10^{12}$ (g), and $1\times10^{13}$ cm$^{-2}$ (h) in Sample B. Background spectra measured in non-implanted regions are also shown.}
\end{flushleft}
\end{figure}

\begin{figure}
\includegraphics[width=7.5truecm]{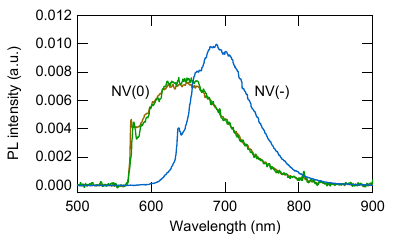}
\end{figure}
\begin{figure}
\begin{flushleft}
{\footnotesize Figure S5 PL spectra of NV$^{0}$ and NV$^{-}$ reference samples. The blue curve is the spectrum of the NV$^{-}$ reference sample (Sample R$_{\mathrm{NV}(-)}$). The green and brown curves are the spectra of the NV$^{0}$ reference samples (green: Sample R$_{\mathrm{NV}(0)}^\mathrm{I}$, brown: Sample R$_{\mathrm{NV}(0)}^\mathrm{II}$). The spectra of the NV$^{0}$ reference samples possibly contain the PL from NV$^{-}$ centers. (See Section S3.)}
\end{flushleft}
\end{figure}

\begin{figure}
\includegraphics[width=16truecm]{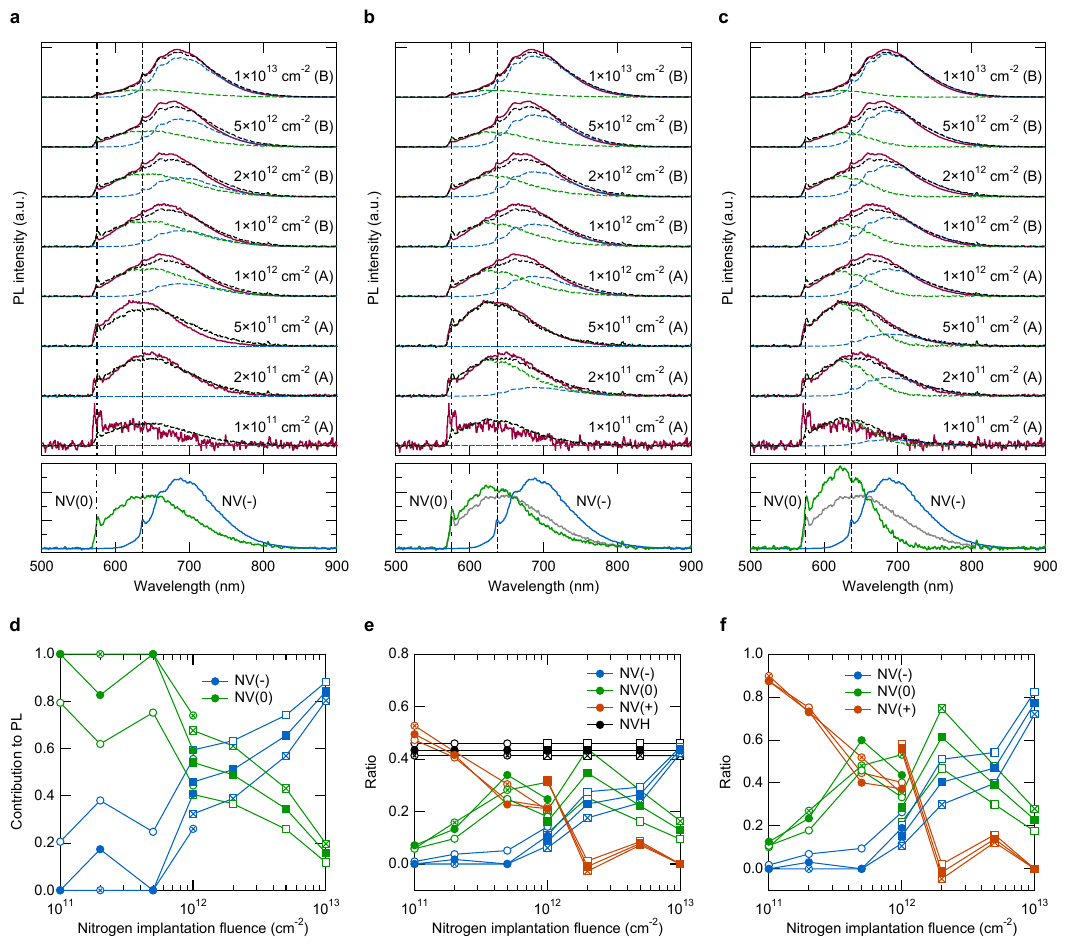}
\end{figure}
\begin{figure}
\begin{flushleft}
{\footnotesize Figure S6 \textbf{a-c} Fitting of the measured spectra (brown curves) with the NV$^{0}$ and NV$^{-}$ reference spectra for cases in which 0\% (\textbf{a}), 20\% (\textbf{b}) or 40\% (\textbf{c}) of the PL of Sample R$_{\mathrm{NV}(0)}^\mathrm{I}$ comes from NV$^{-}$ centers. (See Section S3.) The green and blue dashed curves are the contributions from NV$^{0}$ and NV$^{-}$ centers and the black dashed curves are the sum of them. The vertical dashed dot (dashed) lines indicate the 575-nm (637-nm) zero phonon line of the NV$^{0}$ (NV$^{-}$) centers. The green and blue curves in the bottom panels are the NV$^{0}$ and NV$^{-}$ reference spectra. The gray curve is the spectrum of Sample R$_{\mathrm{NV}(0)}^\mathrm{I}$. \textbf{d} Relative contributions of NV$^{0}$ and NV$^{-}$ to the PL spectra obtained by the fitting. \textbf{e} NVH/NV$^{+}$/NV$^{0}$/NV$^{-}$ relative populations calculated with the data shown in Figs. 4b and S6d. (See Section S4 for the calculation.) \textbf{f} NV$^{+}$/NV$^{0}$/NV$^{-}$ relative populations. In \textbf{d-f}, symbols with crosses, filled symbols, and empty symbols correspond to the 0\% (\textbf{a}), 20\% (\textbf{b}) and 40\% cases (\textbf{c}). Circles and squares represent Samples A and B.}
\end{flushleft}
\end{figure}

\begin{figure}
\includegraphics[width=16truecm]{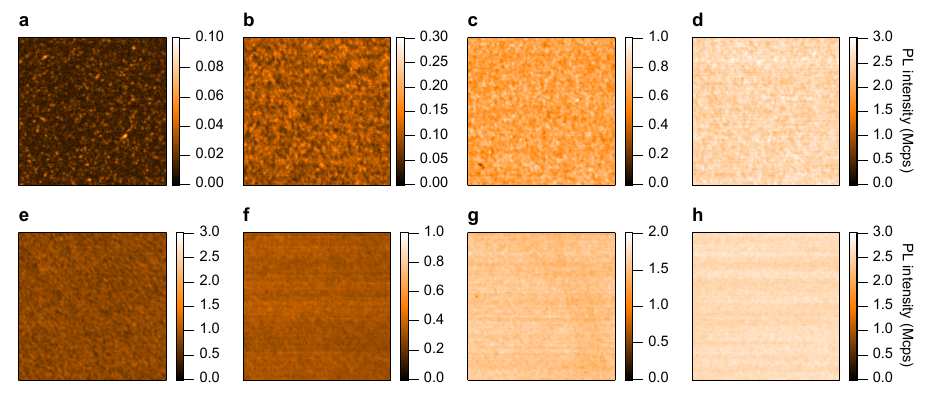}
\begin{flushleft}
{\footnotesize Figure S7 \textbf{a-d} Photoluminescence images for nitrogen implantation fluences of $1\times10^{11}$ (\textbf{a}), $2\times10^{11}$ (\textbf{b}), $5\times10^{11}$ (\textbf{c}), and $1\times10^{12}$ cm$^{-2}$ (\textbf{d}) in Sample A. \textbf{e-h} Photoluminescence images for nitrogen implantation fluences of $1\times10^{12}$ (\textbf{e}), $2\times10^{12}$ (\textbf{f}), $5\times10^{12}$ (\textbf{g}), and $1\times10^{13}$ cm$^{-2}$ (\textbf{h}) in Sample B. The image sizes are $20$ $\mu$m $\times$ $20$ $\mu$m. The excitation laser intensity is 20 $\mu$W for an implantation fluence of ${\ge} 2\times10^{12}$ cm$^{-2}$ and 200 $\mu$W for lower implantation fluences.}
\end{flushleft}
\end{figure}

\newpage

\begin{figure}
\includegraphics[width=12truecm]{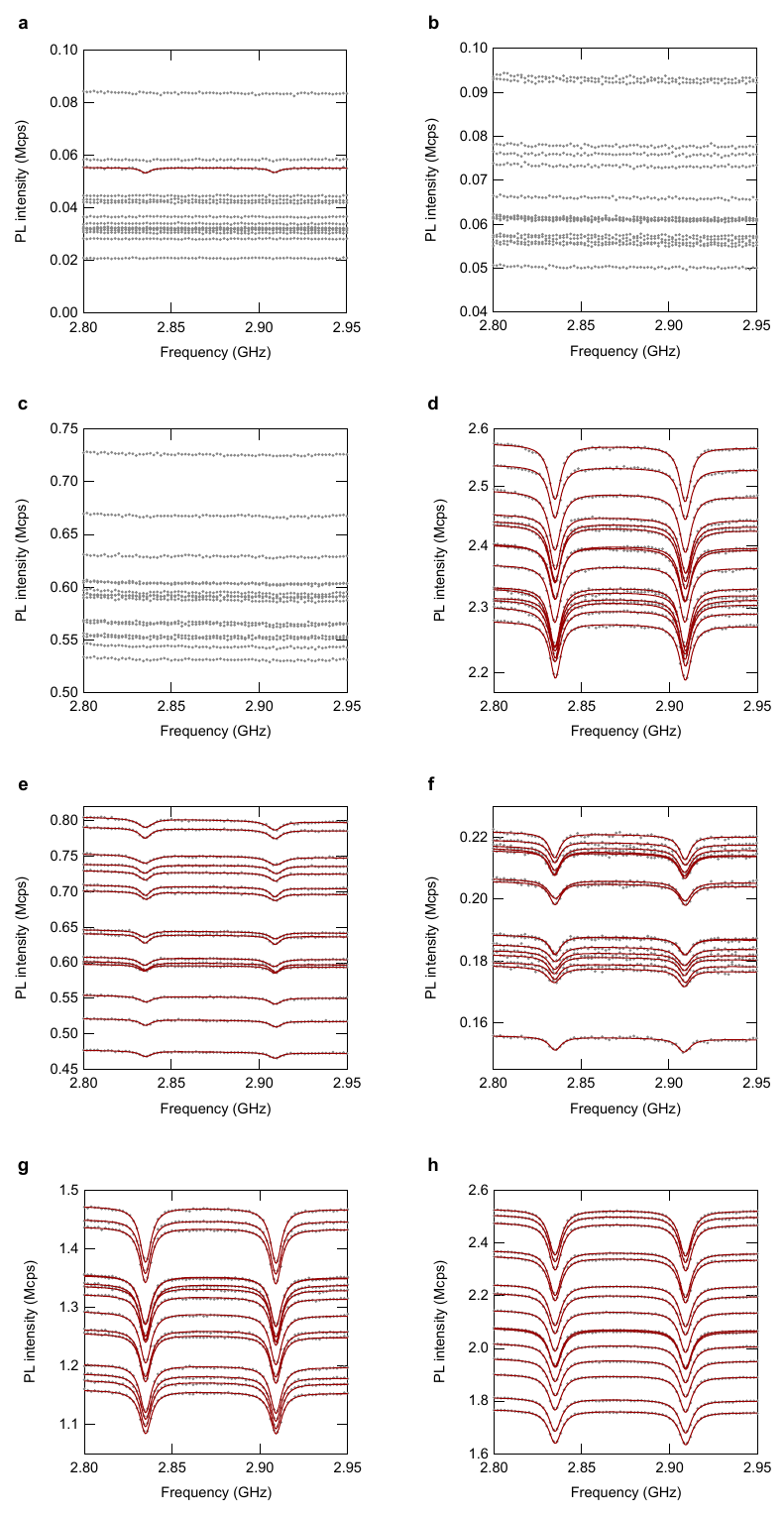}
\end{figure}
\newpage
\begin{figure}
\begin{flushleft}
{\footnotesize Figure S8: \textbf{a-d} Double Lorentzian fits (lines) to ODMR spectra (gray dots) in the regions of nitrogen implantation fluences of $1\times10^{11}$ (\textbf{a}), $2\times10^{11}$ (\textbf{b}), $5\times10^{11}$ (\textbf{c}), and $1\times10^{12}$ cm$^{-2}$ (\textbf{d}) in Sample A. \textbf{e-h} Double Lorentzian fits (lines) to ODMR spectra (gray dots) in the regions of nitrogen implantation fluences of $1\times10^{12}$ (\textbf{e}), $2\times10^{12}$ (\textbf{f}), $5\times10^{12}$ (\textbf{g}), and $1\times10^{13}$ cm$^{-2}$ (\textbf{h}) in Sample B. The ODMR spectra are the same as those shown in Fig. 6. The excitation laser intensity is 20 $\mu$W for an implantation fluence of ${\ge} 2\times10^{12}$ cm$^{-2}$ and 200 $\mu$W for lower implantation fluences. A linear background is assumed for the fitting.}
\end{flushleft}
\end{figure}

\begin{figure}
\includegraphics[width=7.5truecm]{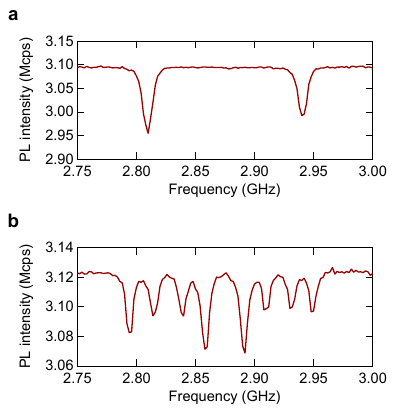}
\begin{flushleft}
{\footnotesize Figure S9: ODMR spectra observed for different magnetic field orientations. \textbf{a} Two ODMR dips appear, indicating that the magnetic field applied with a permanent magnet is parallel to the [001] direction of diamond. \textbf{b} Eight ODMR dips corresponding to the four axes of NV centers appear. The permanent magnet direction was tilted from that used for \textbf{a}.}
\end{flushleft}
\end{figure}

\begin{figure}
\includegraphics[width=15.5truecm]{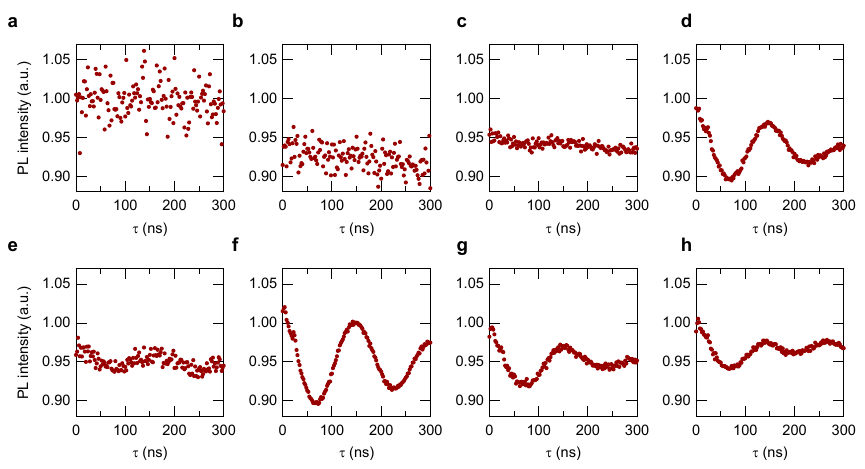}
\begin{flushleft}
{\footnotesize Figure S10 \textbf{a-d} Rabi oscillations measured in regions with nitrogen implantation fluences of $1\times10^{11}$ (\textbf{a}), $2\times10^{11}$ (\textbf{b}), $5\times10^{11}$ (\textbf{c}), and $1\times10^{12}$ cm$^{-2}$ (\textbf{d}) in Sample A. \textbf{e-h} Rabi oscillations in regions with nitrogen implantation fluences of $1\times10^{12}$ (\textbf{e}), $2\times10^{12}$ (\textbf{f}), $5\times10^{12}$ (\textbf{g}), and $1\times10^{13}$ cm$^{-2}$ (\textbf{h}) in Sample B. The excitation laser intensity is 20 $\mu$W for an implantation fluence of ${\ge} 5\times10^{12}$ cm$^{-2}$ and 200 $\mu$W for lower implantation fluences.}
\end{flushleft}
\end{figure}

\begin{figure}
\includegraphics[width=15truecm]{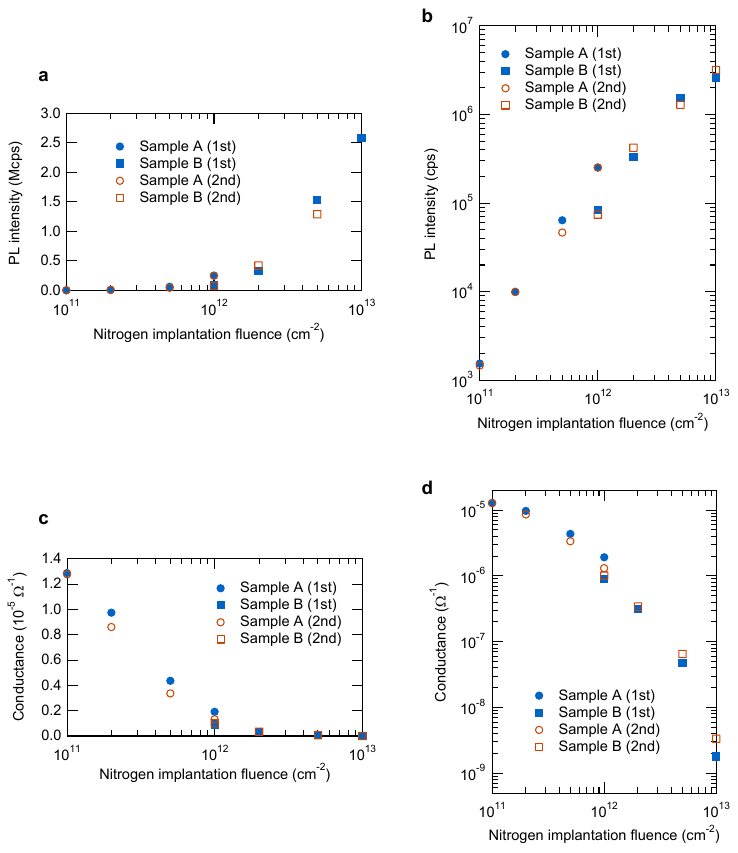}
\end{figure}
\begin{figure}
\begin{flushleft}
{\footnotesize Figure S11 \textbf{a,b} Nitrogen-implantation-fluence dependence of PL intensity measured on different dates. The first measurements on Sample A (B) were carried out 2-3 (2-4) days after the hydrogenation. The second measurements were carried out 233-234 (232-233) days after the hydrogenation. \textbf{c,d} Nitrogen-implantation-fluence dependence of the conductance measured on different dates. The first measurements on Sample A (B) were carried out 6 (7) days after the hydrogenation. The second measurements were carried out 243 (244) days after the hydrogenation.}
\end{flushleft}
\end{figure}

\end{document}